\documentclass{amsart}  
\usepackage{times,amssymb,amsmath,amsfonts,float,nicefrac} 
\usepackage{euscript,graphics}
\usepackage[dvips]{epsfig}
     
\newtheorem{theorem}{Theorem}[section]

\newtheorem{lemma}[theorem]{Lemma}

\renewenvironment{proof}[1]{\noindent {\it Proof~:} #1}
{\ \rule{1mm}{2mm}\medskip}

\renewcommand\qed{\rule{1mm}{2mm}\medskip}

\newcommand{\remove}[1]{}
\renewcommand{\tilde}{\widetilde}

\newcommand\nd{\noindent}

\newcommand\wt{\mbox{{\rm wt}\,}}

\def\shape{\qopname\relax{no}{shape}}
\def\wt{\qopname\relax{no}{w}}

\newcommand\nc\newcommand
\nc\bfa{{\boldsymbol a}}\nc\bfA{{\bf A}}\nc\cA{{\mathcal A}}
\nc\bfb{{\boldsymbol b}}\nc\bfB{{\bf B}}\nc\cB{{\mathcal B}}
\nc\bfc{{\boldsymbol c}}\nc\bfC{{\bf C}}\nc\cC{{\mathcal C}}
\nc\bfd{{\boldsymbol d}}\nc\bfD{{\bf D}}\nc\cD{{\mathcal D}}
\nc\bfe{{\boldsymbol e}}\nc\bfE{{\bf E}}\nc\cE{{\mathcal E}}
\nc\bff{{\boldsymbol f}}\nc\bfF{{\bf F}}\nc\cF{{\mathcal F}}
\nc\bfg{{\boldsymbol g}}\nc\bfG{{\bf G}}\nc\cG{{\mathcal G}}
\nc\bfh{{\boldsymbol h}}\nc\bfH{{\bf H}}\nc\cH{{\mathcal H}}
\nc\bfi{{\boldsymbol i}}\nc\bfI{{\bf I}}\nc\cI{{\mathcal I}}
\nc\bfj{{\boldsymbol j}}\nc\bfJ{{\bf J}}\nc\cJ{{\mathcal J}}
\nc\bfk{{\boldsymbol k}}\nc\bfK{{\bf K}}\nc\cK{{\mathcal K}}
\nc\bfl{{\boldsymbol l}}\nc\bfL{{\bf L}}\nc\cL{{\mathcal L}}
\nc\bfm{{\boldsymbol m}}\nc\bfM{{\bf M}}\nc\cM{{\mathcal M}}
\nc\bfn{{\boldsymbol n}}\nc\bfN{{\bf N}}\nc\cN{{\mathcal N}}
\nc\bfo{{\boldsymbol o}}\nc\bfO{{\bf O}}\nc\cO{{\mathcal O}}
\nc\bfp{{\boldsymbol p}}\nc\bfP{{\bf P}}\nc\cP{{\mathcal P}}
\nc\bfq{{\boldsymbol q}}\nc\bfQ{{\bf Q}}\nc\cQ{{\mathcal Q}}
\nc\bfr{{\boldsymbol r}}\nc\bfR{{\bf R}}\nc\cR{{\mathcal R}}
\nc\bfs{{\boldsymbol s}}\nc\bfS{{\bf S}}\nc\cS{{\mathcal S}}
\nc\bft{{\boldsymbol t}}\nc\bfT{{\bf T}}\nc\cT{{\mathcal T}}
\nc\bfu{{\boldsymbol u}}\nc\bfU{{\bf U}}\nc\cU{{\mathcal U}}
\nc\bfv{{\boldsymbol v}}\nc\bfV{{\bf V}}\nc\cV{{\mathcal V}}
\nc\bfw{{\boldsymbol w}}\nc\bfW{{\bf W}}\nc\cW{{\mathcal W}}
\nc\bfx{{\boldsymbol x}}\nc\bfX{{\bf X}}\nc\cX{{\mathcal X}}
\nc\bfy{{\boldsymbol y}}\nc\bfY{{\bf Y}}\nc\cY{{\mathcal Y}}
\nc\bfz{{\boldsymbol z}}\nc\bfZ{{\bf Z}}\nc\cZ{{\mathcal Z}}
\nc\od{{\bar d}}\nc\ow{{\bar w}}\nc\odelta{{\bar\delta}}
\nc\ox{{\bar x}}\nc\oy{{\bar y}}\nc\ou{{\bar u}}
\nc\oh{{\bar h}}

\newcommand\reals{{\mathbb R}}
\newcommand\complexes{{\mathbb C}}
\newcommand\ff{{\mathbb F}}

\newcommand\integers{{\mathbb Z}}
\nc\dgv{\delta_{\text{\rm GV}}}
\nc\dcrit{\delta_{\text{\rm{crit}}}}
\nc\Esp{E_{\text{\rm sp}}}
\renewcommand\epsilon{\varepsilon}

\nc\hr{\overrightarrow{H}}
\nc\hl{\overleftarrow{H}}

\newcommand{\beeq}{\begin{eqnarray*}}
\newcommand{\eneq}{\end{eqnarray*}}

\newcommand{\half}{\nicefrac12}

\begin{document}
\title[Ordered codes and arrays]	
{Bounds on ordered codes and orthogonal arrays} \thanks{{\em Date}\/: \today.\/ {\em AMS Subject Classification}: Primary 05E30, Secondary 94B65. Research 
supported in part by NSF grants
CCF0515124 and CCF0635271, and by NSA grant H98230-06-1-0044.}%

\author[A. Barg]{Alexander Barg$^\ast$}\thanks{$^\ast$
Dept. of Electrical and Computer Engineering and Institute for Systems 
Research, University of Maryland, College Park, MD 20742, USA, 
and Institute for Problems of Information Transmission, Russian Academy of Sciences, Moscow, Russia. Email: abarg@umd.edu.}
\author[P. Purkayastha]{Punarbasu Purkayastha$^\dag$}\thanks{$^\dag$ 
Dept. of Electrical and Computer Engineering, University of Maryland,
College Park, MD 20742. Email: ppurka@umd.edu.}

\begin{abstract}
We derive new estimates of the size of codes and orthogonal arrays
in the ordered Hamming space (the Niederreiter-Rosenbloom-Tsfasman space). 
We also show that the eigenvalues of the ordered Hamming scheme,
the association scheme that describes the combinatorics of the space,
are given by the multivariate Krawtchouk polynomials, and establish some
of their properties.
\end{abstract}

\maketitle
{\small \tableofcontents}
\section{Introduction}
\subsection{The Niederreiter-Rosenbloom-Tsfasman metric space}

Let $\cQ$ be a finite alphabet of size $q$ viewed as an additive group
mod $q$. Consider the set $\cQ^{r,n}$ of vectors
of dimension $rn$ over $\cQ$. A vector $\bfx$ will be written as
a concatenation of $n$ blocks of length $r$ each, 
$\bfx=\{x_{11},\dots,x_{1r};\dots;x_{n1},\dots,x_{nr}\}.$ 
 For a given vector $\bfx$ let $e_i, i=1,\dots,r$ be the number of $r$-blocks
of $\bfx$ whose rightmost nonzero entry is in the $i$th position counting
from the beginning of the block. The $r$-vector $e=(e_1,\dots,e_r)$ will
be called the {\em shape} of $\bfx$. 
 For two vectors $\bfx,\bfy\in \cQ^{r,n}$ let us write $\bfx \sim_e \bfy$ if 
$\shape(\bfx-\bfy)=e$.
A shape vector $e=(e_1,\dots,e_r)$ 
defines a partition 
of a number $N\le n$ into a sum of $r$
parts. 
Let 
$\Delta_{r,n}=\{e\in (\integers_+\cup\{0\})^r: \sum_ie_i\le n\}$ 
be the set of all such partitions.
 For brevity we write
  $$
     |e|=\sum_i e_i, \quad |e|'=\sum_{i} ie_i, \quad e_0=n-|e|.
  $$

Let $\bfx\in \cQ^{r,n}$ be a vector of shape $e$. Define a weight
function (norm) on $\cQ^{r,n}$
by setting $\wt(\bfx)=|e|'$ and let $d_r(\bfx,\bfy)=\wt(\bfx-\bfy)$
denote the metric induced by this norm.
We call the function $d_r$ the {\em ordered weight}.
The ordered weight
was first introduced by Niederreiter \cite{nie86} in his study 
of low-discre\-pancy point sets. Later, Rosenbloom and Tsfas\-man \cite{ros97} 
independently defined the weight $\wt(\bfx)$ (more precisely, the 
weight $\bar \wt$ defined below), calling it the $m$-metric, 
and studied codes in $\cQ^{r,n}$ with respect to it. 
The set $\cQ^{r,n}$ together with this metric will be called the {\em ordered
Hamming space} (the {\em NRT space}) and denoted by $\hr=\hr(q,n,r)$. 
Unless specified otherwise, below by distance (weight) we 
mean the ordered distance (weight) for some fixed value of 
$r$.
Note that the case $r=1$ corresponds to the usual Hamming distance 
on $\cQ^n$. 

An $(nr,M,d)$ {\em ordered code} $C\subset \cQ^{r,n}$ is an arbitrary subset of
$M$ vectors in $\cQ^{r,n}$ such that the ordered distance between any
two distinct vectors in $C$ is $d$ or more. If $\cQ$ is a finite field
and $C$ is a linear code of dimension $k$, we refer to it as an $[nr,k,d]$
code. 

To define ordered orthogonal arrays,
let us call a subset of coordinates $\cI\subset \{1,\dots,rn\}$
{\em left-adjusted} if with any coordinate $ir+j,0\le i\le n-1,1\le j\le r$ 
in the $i$th block it also contains all the
coordinates $(ir+1,\dots,ir+j-1)$ of the same block.
A subset $A\subset \cQ^{r,n}, |A|=M$ is called a $(t,n,r,q)$ 
{\em ordered orthogonal array} (OOA)
of strength $t$ if its projection on any left-adjusted set of $t$ coordinates
contains all the $q^t$ rows an equal number, say $\theta$, of times.
The parameter $\theta$ is called the {\em index} of $A$.
It follows that $M=\theta q^t$. If $C$ is a linear $[nr,k,d]$ code then
the code orthogonal to it with respect to the dot product is a $(d-1,n,r,q)$ 
linear OOA of index $\theta=q^{nr-k-d+1}.$ 
If $\cQ$ is equipped with the structure of
an additive group, then one can construct {\em additive} OOAs. 
If $\cQ$ is a finite field, it is possible to construct linear OOAs.

OOAs (also called hypercubic designs) were 
introduced in Lawrence \cite{law96} and Mullen and Schmid \cite{mul96}
as a combinatorial equivalent of point sets suitable for numerical integration
over the cube. Informally this link can be described as follows. 
Let $C_n=[0,1]^n$ be a unit cube, let $f$ be a continuous
function of bounded variation and let $\cN$ be a set of $M=q^m$ points
in $C_n,$ called a net. It is known that the error of quasi-Monte Carlo
integration
  $
    |\int_{C_n} f dx-q^{-m}\sum_{x\in \cN} f(x)|
  $
can be bounded above by $V(f)D(\cN)$ where $V(f)$ is 
the total variation of $f$ on $C_n$ and $D(\cN)$ is 
the discrepancy factor  of the net. The parameter $D(\cN)$ measures
the deviation of the net from a uniformly distributed set of points.
The study of uniformly distributed point sets was initiated by
 H.~Weyl and E.~Hlawka. $\text{Sobol}$ \cite{sob67} 
developed the notions described above and
gave the first constructions of nets with bounded factor $D.$
The study of nearly uniform point sets was taken up in Niederreiter
\cite{nie86} which put forward the notion of $(t,m,s)$-nets and derived
a bound on $D(\cN)$ via its parameters. 
We refer to \cite{mar06,nie01a,nie04} for detailed background and more
references for $(t,m,s)$-nets and in particular, to the literature
on their constructions.

The following theorem relates OOAs and $(t,m,s)$-nets. Below we use its
statement as a definition of a $(t,m,s)$-net.
\begin{theorem}\label{thm:tms}
{\rm (Lawrence \cite{law96}, Mullen and Schmid \cite{mul96}).} 
There exists a $(t,m,s)$-net
if and only if there exists an $(m-t,s,m-t,q)$ OOA of index $q^t$ and
size $M=q^m.$
\end{theorem}

Independently of this line of work, Rosenbloom and Tsfasman \cite{ros97}
considered codes in the ordered Hamming space and derived bounds on 
their size.
It became clear shortly after their work that OOAs and
ordered codes are dual types of objects in the sense of Delsarte's algebraic 
theory of coding.
This link opened up an avenue for
applications of coding-theoretic methods to the study of $(t,m,s)$-nets
and motivated the study of ordered codes and OOAs independently of these
applications. In particular, Martin \cite{mar99,mar00a,mar00b,mar07}
has constructed an association scheme that describes the combinatorics of 
the space $\hr$, formulated the linear programming bound (LP bound), outlined a
construction of orthogonal polynomials that describe the eigenvalues of the
scheme and derived Plotkin and Rao bounds on OOAs via linear programming.
Much attention was also devoted to relations between the weight enumerators
of linear codes and their duals in the ordered Hamming space and its
generalizations. In particular, a MacWilliams theorem for the NRT space was 
derived by Martin \cite{mar99}, see also Skriganov \cite{skr01}, 
Dougherty and Skriganov \cite{dou02}.
Bierbrauer \cite{bie02,bie05a,bie06a} studied coding 
constructions of $(t,m,s)$-nets and bounds for them including the LP bound. 
Apart from the combinatorial context, ordered codes arise 
in a number of applied problems such as recent algebraic list decoding 
algorithms of Reed-Solomon codes \cite{nie01b}, a study of linear
complexity of sequences \cite{mas96}, and in a problem in communication
theory \cite{gan07}.

In this paper, we derive several new bounds on OOAs and ordered codes.
We begin with a Bassalygo-Elias bound on codes which improves the known
upper bounds on their size. However, the bulk of our results are devoted 
to code
bounds using the approach via association schemes and linear programming.
We begin with a study of the eigenvalues of the ordered Hamming scheme,
an association scheme that describes the combinatorics of the space
$\hr.$ The approach that we follow relies on the orthogonality relation
of the eigenvalues. This enables us to identify the eigenvalues as
multivariate generalizations of the well-known Krawtchouk polynomials, i.e.,
a family of real polynomials of $r$ discrete variables orthogonal on
$\Delta_{r,n}$ with respect to the weight given by the multinomial
probability distribution. This is the subject of Section~\ref{sect:Ac}.
Turning to bounds, in Sect.~\ref{sect:lpb} 
we derive a new universal estimate of the
size of ordered codes and OOAs with a given distance (strength).
The asymptotic version of this estimate improves the other results in a
certain range of rates. The final section is devoted to the
case $r=2$ for which the bounds can be further improved relying on a 
direct approach.

\subsection{Notation}

Together with the space $\hr$ we will consider the space $\hl(q,n,r)$ 
which differs 
from it in that the vectors are read ``from right to left.'' Namely,
for $\bfx\in \cQ^{r,n}$ let $
   \overline{\shape}(\bfx)=(e_1,\dots, e_r),
  $
where $e_j$ is the number of blocks $(x_{i1},\dots,x_{ir})$ such that
$x_{i1}=\dots=x_{i,r-j}=0$ and $x_{i,r-j+1}\ne 0.$ Let $\bar \wt(\bfx)=|e|',$
where $e=\overline{\shape}(\bfx),$ and let $\bar d_r(\bfx,\bfy)=
\bar\wt(\bfx-\bfy).$

These metric spaces are identical to each other; the reason for considering
them both is that if we equip $\hr$ with the structure of an association
scheme then its dual scheme in Delsarte's sense gives rise to the space
$\hl.$ In particular, if $\cQ$ is a finite field and $C$ is a linear code
in $\hr$ then its dual code $C^\bot$ lives in $\hl.$ We elaborate on this 
below.

An easy combinatorial calculation shows that
the number of vectors of shape $e$ in $\hr$ is given
by
  \begin{equation}\label{eq:v}
    v_e=\binom n{e_0,e_1,\dots,e_r} ({q-1})^{|e|} q^{|e|'-|e|}
  \end{equation}
and the number of vectors of weight $d$ equals to
  \begin{equation}
    S_d=\sum_{e: |e|'=d} v_e.
  \end{equation}
Let $A(z)=(q-1)z(z^r-1)/(q(z-1))$ and let $z_0=z_0(x)$ satisfy
the equation
   \begin{equation}\label{eq:z0}
     x r(1+A(z))= \frac{q-1}q\sum_{i=1}^r iz^i.
   \end{equation}
Define the function 
 $$
     H_{q,r}(x)=x(1-\log_q z_0)+\frac 1r\log_q(1+A(z_0)).
  $$
In the case $r=1$ we write $h_q(x)$ instead of $H_{q,1}(x)$, where
   $$
     h_q(x)=- x\log_q\frac x{q-1}-(1-x)\log_q(1-x).
   $$
Let \begin{equation}\label{eq:dcrit}
  \dcrit=1-\frac 1r\sum_{i=1}^r q^{-i}=1-\frac 1{rq^r}\frac{q^r-1}{q-1}.
   \end{equation}
The asymptotic volume of the sphere in $\hr$ is
given in the next lemma.
\begin{lemma} \label{lemma:vol} {\rm \cite{ros97}}
  (a) For $0<x<1$, equation (\ref{eq:z0}) has a unique positive
root $z_0(x), z_0\in[0,r].$\\
  (b) Let $r\ge 1$ be fixed, $n\to\infty, d/n\to r\delta,$ then
         \begin{equation}\label{eq:sphere}
    \lim_{n\to\infty}(nr)^{-1}\log_q\sum_{i=0}^d S_i=
   \begin{cases} H_{q,r}(\delta), &0\le \delta\le\dcrit,\\
   1, & \dcrit<\delta\le 1.
  \end{cases}
  \end{equation}
\end{lemma}

\section{Bounds on Ordered Codes and Arrays}

In general, given the value of the distance or the strength, 
our goal is to construct as large codes and as small OOAs as possible.
The latter will also account for small-size $(t,m,s)$-nets with bounded discrepancy.
In this section we recall the known bounds on ordered codes and OOAs and 
derive a new bound on the size of codes. 

\subsection{Existence bounds}
A Gilbert-type bound on ordered codes was derived in \cite{ros97}.
\begin{theorem} There exists an $(nr,M,d)$ code in the space
$\hr$ whose parameters satisfy
    $$
      M\sum_{i=0}^{d-1}S_i\ge q^{nr}.
    $$
If $\cQ$ is a finite field, then there exists a linear code with the same
parameters. 
\end{theorem}
A bound that applies specifically to linear codes was proved in 
\cite{bie02}. It is analogous to the Varshamov bound for the Hamming space.
\begin{theorem} \label{thm:var} 
Suppose that $m$ and $t$ satisfy the conditions
  $$
    \sum_{i=0}^{t-\tau} S_{i,n-1}< q^{m-(\tau-1)}, \quad
         \tau=1,\dots,t-1.
  $$
Then there exists an $[nr,nr-m]$ linear code in $\hr$ of distance $\ge t+1$,
and a $(t,n,r,q)$ linear OOA of dimension $m$.
\end{theorem}

\subsection{Nonexistence bounds}
While in general bounds on codes do not imply lower bounds on 
OOAs, there are two special cases when these two types of results are
equivalent. First, if $C$ is an $[nr,k,d]$ linear code over $\ff_q^{nr}$ 
then the code $C':=\{\bfy\in \ff_q^{nr}: \sum_{i=1}^{nr}
x_iy_i=0 \mbox{ for all } {\bfx\in C}\}$ is a $(d-1,n,r,q)$ linear OOA. 
Next, if an upper (lower) bound on codes (OOAs) is obtained by linear
programming as explained in the next section then the same solution
of the LP problem gives a lower (upper) bound on OOAs (codes).

We next mention some upper bounds on codes and OOAs.

\subsubsection {Singleton bound} The parameters of any $(nr,M,d)$ code satisfy
   $$
     M\le q^{nr-d+1}.
   $$

\subsubsection{Plotkin bound} A {Plotkin bound} on codes was established 
in \cite{ros97}. Namely, the following result holds true.
\begin{theorem}\label{thm:plotkin} Let $C\subset \hr$ be a code of size $M$ and 
distance $d>nr\dcrit$.
Then
    $$
        M\le \frac {d}{d-nr\dcrit}.
    $$
\end{theorem}

A dual Plotkin bound on OOAs was proved by Martin and Visentin \cite{mar07}.
\begin{theorem} \cite{mar07}
     Let $C$ be a $[t,n,r,q]$ OOA. If $t> nr\dcrit-1$ then
      $$
       |C|\ge q^{nr}\Big(1-\frac{nr\dcrit}{t+1}\Big).
      $$
\end{theorem}

\subsubsection{Hamming-Rao bound} According to the Hamming bound, 
the parameters of any $(nr,M,d=2\tau+1)$ code
satisfy
         $$
            M \le \frac{q^{rn}}
            {\sum_{i=0}^\tau S_i}.
         $$
A dual bound in this case is the Rao bound which for the NRT space
was established by Martin and Stinson \cite{mar00a}: The size $M$
of any $(t=2\tau,n,r,q)$ OOA satisfies
       $$
            M\ge \sum_{i=0}^\tau S_i.
       $$

\subsubsection{A Bassalygo-Elias bound on codes} The next result is new.

\begin{theorem}\label{thm:be} Let $C$ be an $(nr,M,d)$ code. Then

        $$
      M\le q^{rn} dn
      \frac{1}{S_w(dn-2wn+\frac{w^2}{r\delta_{\text{crit}}})}.
           $$
for any $  w\le nr \dcrit(1-\sqrt{1-d/(nr\dcrit)}).$
\end{theorem}
\begin{proof} We will rely upon the next lemma.
{\begin{lemma}\label{lemma:johnson}
  Let $C\subset \hr, |C|=M$ be a code all of whose vectors have weight $w$
  and are at least distance $d$ apart.
  Then for $d\ge  2w-w^2/(nr\dcrit)$,
     $$
       M\le \frac{dn}{dn-2wn+\frac{w^2}{r\delta_{\text{crit}}}}.
     $$
\end{lemma}
\begin{proof} 
Let $C^i$ be a projection of $C$ on the $i$th block
of coordinates. For a vector $\bfz\in \cQ^r$ let
$\bfz^h=(z_{r-h+1},\dots,z_r)$ be its suffix of length $h$. 
Given $\bfx\in C$, we denote by $\bfx^i\in C^i$ its $i$th block
and write $\bfx^{i,h}$ to refer to the $h$-suffix of $\bfx^i.$
 For
$i=1,\dots,n; h=1,\dots,r; \bfc\in \cQ^h$ let
  $$
    \lambda_{i,\bfc}^h=|\{\bfx^i\in C^i:\; \bfx^{i,h}=\bfc\}|
  $$
be the number of vectors in the $i$th block whose $h$-suffix equals $\bfc.$
We have
   \begin{align*}
     d_r(\bfx^i,\bfy^i)&=r-\sum_{h=1}^r \delta(\bfx^{i,h},
         \bfy^{i,h})\\&=r-\sum_{h=1}^r\sum_{\bfc\in \cQ^h}
            \delta(\bfx^{i,h},\bfc)\delta(\bfy^{i,h},\bfc).
  \end{align*}
Compute the sum of all distances in the code as follows:
 \begin{align}
    \sum_{\bfx,\bfy\in C}d_r(\bfx,\bfy)
          &=nrM^2-\sum_{i=1}^n\sum_{\bfx^i,\bfy^i\in C^i}
         \sum_{h=1}^r\sum_{\bfc\in \cQ^h}
            \delta(\bfx^{i,h},\bfc)\delta(\bfy^{i,h},\bfc)\nonumber\\
         &=nrM^2-
\sum_{i=1}^n\sum_{h=1}^r\sum_{\bfc\in \cQ^h}
          (\lambda_{i,\bfc}^h)^2.\label{eq:sum1}
  \end{align}
To bound above the right-hand side, we need to find the minimum of the 
quadratic form
   $$
    F=\sum_{i=1}^n\sum_{h=1}^r\sum_{c\in \cQ^h\backslash\{0\}}
    (\lambda_{i,c}^h)^2+\sum_{i=1}^n\sum_{h=1}^r
    (\lambda_{i,0}^h)^2
   $$
under the constraints
  \begin{equation}\label{eq:hyp}
    \sum_{i=1}^n\sum_{h=1}^r\lambda_{i,0}^h=M(nr-w), \quad 
   \sum_{c\in \cQ^h}\lambda_{i,c}^h=M \;(1\le h\le r, \;1\le i\le n).
  \end{equation}
Critical points of $F$ in the intersection of these hyperplanes,
together with (\ref{eq:hyp}), satisfy the equations
   \begin{equation}\label{eq:lag}
      \begin{array}{l@{\quad}l}
     2\lambda_{i,c}^h+\beta_{i,h}=0,       &1\le i\le n;1\le h\le r;
                     c\in \cQ^h\backslash\{0\}\\[1mm]
     2\lambda_{i,0}^h+\alpha+\beta_{i,h}=0,&1\le i\le n;1\le h\le r\end{array}
    \quad\alpha, \beta_{i,h}\in\mathbb{R}.
   \end{equation}
The system (\ref{eq:hyp})-(\ref{eq:lag}) has a unique solution for the 
variables 
$\lambda_{i,c}^h,\beta_{i,h},\alpha$; in particular, 
   $$
    \lambda_{i,0}^h=M\Big[\Big(\frac 1{q^h}-1\Big)
\frac w{nr\dcrit}+1\Big], \quad h=1,\dots,r, i=1,\dots,n,
   $$
  $$
   \lambda_{i,c}^h=\frac{Mw}{q^h nr\delta_{\text{crit}}}, \quad
           \quad h=1,\dots,r, i=1,\dots,n, c\in \cQ^h\backslash\{0\}.
  $$
To verify that this critical point is in fact a minimum, observe that
the form $F$ is convex because its Hessian matrix is $2I$ and is 
positive definite
(both globally and restricted to the intersection of the hyperplanes 
(\ref{eq:hyp}) ).
Substituting these values of the $\lambda$s 
and taking account of the fact that $\sum_h q^{-h}=r(1-\delta_{\text{crit}}),$
we get 
  $$
     F\ge \sum_{i=1}^n\sum_{h=1}^r\sum_{c\ne 0} \Big(\frac{Mw}{q^{h}
       nr\dcrit}\Big)^2 + \sum_i\sum_h M^2\Big[\Big(\frac 1{q^h}-1\Big)
    \frac w{nr\dcrit}+1\Big]^2
  $$
  $$
      = M^2n\Big(\frac{w^2}{n^2r\delta_{\text{crit}}}
           -\frac{2w}{n}+r\Big).
  $$
Then from (\ref{eq:sum1}) we obtain
   $$
     dM(M-1)\le \sum_{\bfx,\bfy\in C} d_r(\bfx,\bfy)\le 
       \frac{M^2}n\Big(2wn-\frac{w^2}
       {r\delta_{\text{crit}}}\Big)
   $$
which gives the result.
\end{proof}}

The proof of the theorem is completed as follows.
     Let $\cS_w\subset \cQ^{r,n}$ be a sphere of radius $w$ around zero.
  Clearly, 
    $$  
    |C||\cS_w|=\sum_{x\in \hr} |(C-x)\cap \cS_w|\le q^{nr} A_q(nr,d,w),
    $$
where $A_q(nr,d,w)$ is the maximum size of a distance-$d$ code in $\cS_w$.
With the previous lemma, this finishes the proof.
\end{proof}

{\em Remarks.}
1. This theorem implies a lower bound on the size $M$ of a linear OOA
$(t-1,n,r,q)$: for any $  w\le nr\dcrit(1-\sqrt{1-t/(nr\dcrit)}),$
   \begin{equation}\label{eq:beooa}
     M\ge \frac 1{tn} S_w\Big(tn-2wn+\frac{w^2}{r\dcrit}\Big)
   \end{equation}
and in particular, a lower bound on linear $(m-r,m,n)$-nets, $m=\log_q M.$

2. Caution should be exercised in dealing with codes 
of a constant weight in the NRT space, 
i.e., codes on the sphere $\cS_w$
in $\hr$. Indeed, the sphere $\cS_w$ together with the metric $d_r$ is not homogeneous: 
in particular, the number of points in $\cS_w$ located at a given distance 
from a point $x\in \cS_w$ depends on $x$. However, this does not cause
problems in the previous theorem.

3. The argument used in the proof of Lemma~\ref{lemma:johnson} can be also 
used to give a proof of the Plotkin bound, 
Theorem \ref{thm:plotkin}, that is simpler than the ones known in the
literature. Indeed, let $C\subset \hr$ be a distance-$d$ code.
Consider again expression (\ref{eq:sum1}) and note that 
this time there is no restriction on the weight of the codewords.
Using the Cauchy-Schwarz inequality and the fact that $\sum_{\bfc\in \ff_q^h}
\lambda_{i,\bfc}^h=M$, we obtain
  $$
    M(M-1)d\le nrM^2-\sum_{i=1}^n\sum_{h=1}^r
        \frac{M^2}{q^h}=nrM^2\dcrit.
  $$
Solving for $M$ concludes the proof. 

\subsubsection{Asymptotics}

In this section we assume that $n\to\infty$ and $r$ is a constant.
 For a code of size $M$ let $R=\log_q M/nr$ be the code {\em rate}.
Given a sequence of $(rn_i,M_i,d_i)$ codes we will say that its
{\em asymptotic rate} is $R$ and the asymptotic relative distance is
$\delta$ if 
   $$
     \lim_{i\to\infty} \frac {1}{rn_i}\log_q M_i=R, \quad \lim_{i\to\infty}
           \frac{d_i}{rn_i}=\delta.
   $$
The Plotkin bound implies that the asymptotic rate and distance of
any sequence of codes satisfy 
   \begin{align*}
     R &\leq 1-\frac\delta\dcrit,  && 0\le \delta\le \dcrit,\\
     R &=0,           &&\delta \ge \dcrit.
   \end{align*}

To state the ``sphere packing'' or ``volume'' bounds on ordered codes
we rely upon Lemma \ref{lemma:vol}. Namely \cite{ros97}, 
there exists a sequence of 
$[rn_i,k_i,d_i]$ linear codes $C_i,i=1,2,\dots,$ such that $n_i\to\infty$,
$k_i/(rn_i)\to R,$ $d_i/(rn_i)\to\delta$ and
   $$
      R\ge 1-H_{q,r}(\delta), \quad 0\le\delta\le \dcrit
    \qquad(\text{Gilbert-Varshamov bound}).
   $$
On the other hand, for any such sequence of codes, 
   $$
      R+H_{q,r}(\nicefrac\delta 2)\le 1 \qquad(\text{Hamming bound}).
   $$

The asymptotic version of Theorem \ref{thm:be} is as follows:
\begin{theorem} {\rm (Asymptotic Bassalygo-Elias bound).} 
 For $0\le \delta\le \dcrit$ the
asymptotic rate and distance of any sequence of codes satisfy
  \begin{equation}\label{eq:be} 
   R\le 1-H_{q,r}\Big(\dcrit(1-\sqrt{1-\delta/\dcrit})\Big).
  \end{equation}
\end{theorem}
This bound is better than the Hamming bound for all $\delta\in(0,\dcrit].$
It is also often better than the Plotkin bound. For instance, for $q=2,r=2$
the bound (\ref{eq:be}) is better than the Plotkin bound for all $\delta
\in(0,\dcrit)$. For larger $q,r$ the improvement is attained only for
low values of $\delta$ since the right-hand side of (\ref{eq:be})
becomes $\cap$-convex close to $\dcrit.$
 For instance, for $q=3,r=4$ this range is $(0,0.54),$ etc.

\subsubsection{Asymptotic bounds for digital $(t,m,s)$-nets} 
A $(t,m,s)$-net is called digital if the OOA that corresponds to it
forms a linear subspace of $\ff_q^{nr}.$ Therefore,
bounds on linear OOAs apply to the special case of digital $(t,m,s)$-nets.
However, studying asymptotics for this case 
requires a different normalization since the
strength $m-t$ of the OOA that corresponds to the net
equals $r,$ and both approach infinity independently of $s$. Therefore, 
let $R=m/s$ denote the rate and 
$\delta=(m-t)/s$ denote the relative strength of the OOA that corresponds
to the net. To state the bounds, we need to compute the asymptotic behavior
of the volume of the sphere, which is different from (\ref{eq:sphere}).
The next result is due to Bierbrauer and Schmid \cite{bie05a}.
  \begin{theorem}\label{thm:gvtms}  
       There exist families of digital $(t,m,s)$-nets with $s,(m-t)\to\infty$
for which $(R,\delta)$ satisfy the bound $R\le \Psi(\delta),$ where
   $$
     \Psi(\delta)=
    \delta-1+\log_q\Big(\frac{q-1+\alpha}\alpha\Big)-\delta\log_q(1-\alpha),
   $$
and $\alpha$ is defined by $\delta\alpha(q-1+\alpha)=(q-1)(1-\alpha).$
  \end{theorem}
On the other hand, by the Rao bound, any family of $(t,m,s)$-nets
satisfies $R\ge \Psi(\delta/2).$ Observe that Theorem \ref{thm:be} in this
case gives the same result as the Rao bound 
because the increase of the packing radius in (\ref{eq:beooa}) over $\delta/2$ 
vanishes asymptotically. Indeed, 
taking $\omega=w/n$ and replacing $t$ with $m-t$, 
we obtain from (\ref{eq:beooa})
  $$
    M\ge \frac 1{\delta} (\delta - 2\omega + o(1)) S_{\omega n}.
  $$
The tightest bound is obtained if we take $\omega=\delta/2$ in this inequality.

{\em Remark:} We note that in the case that 
both $n\to\infty$ and $r\to\infty$ while $\delta=d/nr$ tends to a constant
bounded away from 0 and 1, the lower and upper bounds on codes 
coincide \cite{ros97} (the Gilbert-Varshamov bound converges to the 
Singleton bound). 

\section{Association Schemes, and the Ordered Hamming Scheme}\label{sect:Ac}
The coding-theoretic notion of duality prompted a study of the association
scheme that describes the combinatorics of the NRT space. We briefly recall 
some elements of Delsarte's theory of association schemes \cite{del73}. 
A symmetric association scheme with $D$ classes
is a finite set $X, |X|=N,$ equipped with a set $\cR=\{R_0,R_1,\dots, R_D\}$
of symmetric binary relations  on $X\times X$ such that
  
(i) $R_0=\{(x,x)\}, \; R_i\cap R_j=\emptyset,\; \cup_{i=0}^D R_i=X \times X;$

(ii) For each $0\le i,j,k\le D$ the number
    $$
     p_{i,j}^k=|\{(x,y)\in R_i, (x,z)\in R_j \text{ given that } (y,z)
              \in R_k\}|
    $$
depends only on $(i,j,k).$ Moreover, $p_{i,j}^k=p_{j,i}^k.$

The parameters $p_{i,j}^k$ are called the intersection numbers of the scheme
$\cA=(X,\cR).$ The numbers $v_i=p_{i,i}^0$ are called the valencies of $\cA.$ 
Let $A_i$ be the adjacency matrix of the relation $R_i,i=0,\dots,D.$
It is clear that 
   $$
    A_0=I, \;\sum_{i=0}^D A_i=J  \quad(\text{all-one matrix}),
  $$
and for all $0\le i,j\le D$ the product $A_iA_j$ is contained in the
linear span of $\{A_0,A_1,\linebreak[2]\dots, A_D\}.$ The matrices $A_i$ form a commutative
algebra over $\complexes$ with respect to matrix multiplication, called the
Bose-Mesner algebra. 

It is clear that the Bose-Mesner algebra is also closed under the Hadamard
(elementwise) 
multiplication $\circ$, viz. $A_i\circ A_j=\delta_{i,j}A_j.$
With respect to $\circ,$ this algebra has a basis of primitive idempotents
$\{E_0,E_1,\dots,E_D\}$ that satisfy
  $$
     E_0=\frac 1N J,\quad E_i\circ E_j=\frac 1N \sum_{k=0}^D q_{i,j}^k E_k,
          \;0\le i,j\le D.
 $$
The numbers $q_{i,j}^k$ are nonnegative. They are called the Krein 
parameters of the association scheme.
The quantities $\mu_i=q_{i,i}^0$ are called multiplicities of the 
scheme $\cA.$ The matrices $P$ and $Q$ defined by
   $$
     A_i=\sum_{j=0}^D P_{ji}E_j, \quad 0\le i\le D,
   $$
and 
   $$
     E_j=\frac 1N\sum_{i=0}^D Q_{ij}A_i, \quad 0\le j\le D,
  $$
are called the {\em first} and {\em second eigenvalues} of $\cA$, respectively.
The eigenvalue matrices satisfy $PQ=NI.$ 
 Further, for $0\leq i,j\leq D$ the eigenvalues satisfy
  \begin{equation}\label{eq:flip}
    \mu_i P_{ij}=v_j Q_{ji},
  \end{equation}
  \begin{equation}\label{eq:ort1}
     \sum_{k=0}^D \mu_k P_{ki}P_{kj}=Nv_i\delta_{ij},
  \end{equation}
  \begin{equation}\label{eq:pijk}
     P_{ij}P_{ik}=\sum_{l=0}^D p_{j,k}^l P_{il}, \quad 0\le i\le D.
  \end{equation}
Two $D$-class 
association schemes are called Delsarte duals of each other if the adjacency
matrices $A_i$, the first eigenvalues $P$, and the intersection numbers
$p_{i,j}^k$ of one scheme are, respectively, the idempotents $E_i$, 
the second eigenvalues $Q$ and the Krein numbers $q_{i,j}^k$ of the other.
The duality also exchanges the role of the matrix and Hadamard
multiplication. 
A scheme is called {\em self-dual} if it equals its dual. For instance,
the Hamming scheme 
$H_n=(X=\cQ^n, R_i=\{ (x,y)\in X, d_H(x,y)=i\}, i=0,\dots,n)$
is self-dual. 
One of the manifestations of self-duality in this case, obvious from
the definition, is the MacWilliams theorem that relates 
the weight distribution of an additive code to that of its dual.
A scheme is called {\em formally self-dual} \cite{bro89} if there exists 
some ordering of primitive idempotents under which $P=Q.$ In a formally 
self-dual scheme $v_i=\mu_i$ and $p_{i,j}^k=q_{i,j}^k.$
 Following Delsarte \cite[p.~17]{del73} a scheme $(X^n,\cR)$ is called
an {\em extension} of an $r$-class scheme $\cK=(X,\cD=(D_0,D_1,\dots,D_r))$ 
if its vertex set is the $n$-fold Cartesian product of $X$ and the relations 
$R_e, e\in \Delta_{r,n}$ are given by 
   \begin{align*}
     R_e=\{((x_{11},\dots, x_{1n}),(&x_{21},\dots,x_{2n})):\\
     &|\{j: (x_{1j},x_{2j})\in D_i\}|=e_i,\; i=0,1,\dots,r\}.
   \end{align*}
Apart from \cite{del73} we refer to \cite{ban84,bro89,god93} for the proofs
of these results and more information on association schemes.

The association scheme for the NRT space was constructed and studied
by Martin and Stinson \cite{mar99}. Define an $r$-class 
``kernel scheme'' $\cK(\cQ^{r,1},\cD=(D_0,D_1,\dots,D_r))$ with the relations
   $$
     D_i=\{(\bfx_1,\bfx_2)
\in \cQ^{r,1}\times \cQ^{r,1}: d_r(\bfx_1,\bfx_2)=i\}, \quad i=0,1,\dots, r.
   $$
\remove{   \begin{multline*}
      R_i=\{((x_{11},\dots,x_{1r}),(x_{21},\dots,x_{2r})): x_{1i}\ne
    x_{2i} \text{ if }i>0\\
       \text{ and }x_{1h}= x_{2h},\;h=i+1,\dots,r                \}
  \quad(0\le i\le r).
  \end{multline*}}
\begin{theorem}\label{thm:MS}\cite{mar99} The space $X=\cQ^{r,n}$ together with the relations
  $$
    R_e=\{(\bfx,\bfy)\in X\times X: \bfx \sim_e \bfy\}  \quad
      (e\in \Delta_{r,n})
  $$
forms a formally self-dual association scheme ${\overrightarrow{\cH}}$, 
called the $r$-Hamming
scheme. It can be constructed as an $n$-fold Delsarte extension of $\cK.$

The dual scheme of ${\overrightarrow{\cH}}$ is ${\overleftarrow{\cH}}$ 
whose point set is $X=\cQ^{r,n}$ and the set of relations is given by 
    $$
  R_e=\{(\bfx,\bfy)\in X\times X: \overline{\shape}(\bfx-\bfy)=e\} \quad
      (e\in \Delta_{r,n}).
    $$
\end{theorem}

\section{Multivariate Krawtchouk Polynomials}
In the conventional case of $r=1$, eigenvalues of the Hamming scheme 
are given by the Krawtchouk polynomials 
      \begin{equation}\label{eq:k}
       k_i(n,x)=\sum_{l=0}^i (-1)^l(q-1)^{k-l} \binom x 
\ell \binom{n-x}{k-\ell}
   \end{equation}
 which form a family
of polynomials of one discrete variable orthogonal on the set
$\{0,1,\dots,n\}$ with weight $\alpha(i)=\binom ni (q-1)^iq^{-n},$
i.e., the binomial probability distribution.
Here we are interested in their generalization for the $r$-Hamming scheme.

Observe that the valencies $v_e=p_{e,e}^0$ of the scheme are given by
(\ref{eq:v}). By self-duality and (\ref{eq:ort1}), the eigenvalues
are orthogonal on the space of partitions $\Delta_{r,n}$ with weight $v_e.$
Below it will be convenient to normalize the weight. Let $V=V_{r,n}$ be the 
space of real polynomials of $r$ discrete
variables $x=(x_1,x_2,\dots, x_r)$ defined on 
$\Delta_{r,n}.$ Let us define a bilinear form acting on the space $V$
by 
    \begin{equation}\label{eq:form}
    \langle u_1,u_2\rangle=\sum_{e\in \Delta_{r,n}} u_1(e)u_2(e)w(e),
  \end{equation}
where $w(e)=q^{-nr}v_e.$ 
Letting $p_i=q^{i-r-1}(q-1), i=1,\dots,r;\, p_0=q^{-r},$ we observe that
   $$
     w(e)={n!}\prod_{i=0}^r \frac{p_i^{e_i}}{e_i!}
   $$
forms a multinomial probability distribution on $\Delta_{r,n}$.
Therefore, $r$-variate polynomials orthogonal with respect to this weight
form a particular case of multivariate Krawtchouk polynomials.

 For a partition $f\in \Delta_{r,n}$ denote by 
  $$K_f(x)=K_{f_1,\dots,f_r}(x_1,\dots,x_r)
  $$ 
the Krawtchouk polynomial that corresponds to it. Let $\kappa=|f|$ be
the degree of $K_f.$ 
\remove{
Next we recognize the fact that $r$-variate polynomials on $\Delta_{r,n}$
orthogonal with respect to $w(e)$ form a particular case of 
multivariate Krawtchouk polynomials. 
General multivariate Krawtchouk polynomials \cite{tra89}
are orthogonal with respect to the multinomial probability distribution
   $$
     \Pr(e)={n!}\prod_{i=0}^r \frac{p_i^{e_i}}{e_i!},
 $$
where $p=(p_0,p_1,\dots,p_r)$ is a vector of probabilities and 
$e=(e_1,\dots, e_r)\in \Delta_{r,n}$ is a partition.
 For any fixed set of nonzero probabilities $p,$ these polynomials
form an orthogonal basis of the space $V=L_2(\Delta_{r,n})$ 
of real polynomials of $r$ discrete variables.
The particular case relevant for our application is given by 
   $$
        p_i=q^{i-r-1}(q-1), i=1,\dots,r;\quad p_0=q^{-r}
   $$
since then $\Pr(e)=w(e).$}

Our goal in this section is to derive properties of the polynomials
$K_f.$ In their large part, these properties are obtained by specializing 
to the current case general relations of the previous section.
However, some work is needed to transform them to a concrete form which
will be used in later calculations.

The following relations are useful below.
\begin{lemma}\label{lemma:xixj}
  \begin{align}
\langle x_i,1\rangle&=n(q-1)q^{i-r-1}, \quad
               &i=1,\dots, r\label{eq:x1}\\
       \langle x_i,x_j\rangle&
    =n(n-1)(q-1)^2q^{i+j-2r-2}, &1\le i\ne j\le r\label{eq:xixj}\\
      \langle x_i,x_i\rangle&=n(q-1)q^{i-r-1}(1+(n-1)(q-1)q^{i-r-1}),
          &i=1,\dots, r. \label{eq:xixi}
  \end{align}
        \end{lemma}
\begin{proof}
To prove (\ref{eq:x1}), compute
   $$
       \langle x_i,1\rangle= q^{-nr}\sum_e\Big\{e_i
     \binom{n}{e_0,e_1,\dots,e_r}\prod_{j=1}^r ((q-1)q^{j-1})^{e_j}\Big\}
    $$
   $$
    =nq^{-nr}
    \sum_e \binom{n-1}{e_0,e_1,\dots,e_i-1,\dots,e_r}
      \prod_{j=1}^r ((q-1)q^{j-1})^{e_j}. 
   $$
The sum on $e$ on the last line
equals $(q-1)q^{i-1+(n-1)r}$ which finishes the proof. The remaining
two identities are proved in a similar way.
\end{proof}

\subsection{Properties of the polynomials $K_e$} \hspace*{\fill}
\linebreak
\medskip\hspace*{8pt}
(i) $K_e(x)$ is a polynomial in the variables $x_1,\dots,x_r$ of degree
$\kappa=|e|.$ 
There are $\binom{\kappa+r-1}{r-1}$ different polynomials of the same
degree, each corresponding to a partition of $\kappa$.

\medskip
(ii) ({\em Orthogonality}) Equation (\ref{eq:ort1}) is rewritten as
  \begin{equation}\label{eq:ort}
     \langle K_{f}, K_{g}\rangle=v_f \delta_{f,g}, \quad \|K_f\|=\sqrt {v_f}.
  \end{equation}
In particular, let $F_i=(0^{i-1}10^{r-i-1}),i=1,\dots,r$ be a partition
with one part. We have
   \begin{equation}\label{eq:ortl}
     \|K_{F_i}\|^2=
    \langle K_{F_i},K_{F_i}\rangle= n(q-1)q^{i-1}, \quad i=1,\dots,r.
   \end{equation}

Indeed, equality (\ref{eq:ort}) is simply (\ref{eq:ort1}) specialized 
to the case at hand and (\ref{eq:ortl}) is obtained from (\ref{eq:v}).

(iii) ({\em Linear polynomials}) For $i=1,\dots,r,$
  \begin{equation}\label{eq:lin} 
  K_{F_i}(x)=q^{i-1}(q-1)(n-x_r-\dots-x_{r-i+2})-q^i x_{r-i+1}.
   \end{equation}

{\em Proof:} This is shown by orthogonalizing the set of linear polynomials
$(1,x_1,x_2,\dots,\linebreak[2]x_r)$. We take $K_{0,\cdots,0}=1.$ 
 Use Lemma \ref{lemma:xixj} to compute
 $$
   K_{F_1}(x)=c_1( x_r-\langle x_r ,1\rangle)=c_1(x_r- n(q-1)/q)
 $$
for some constant $c_1$. To find $c_1$, use (\ref{eq:ortl}):
  $$
   n(q-1)=c_1^2 \|x_r-\frac{n(q-1)}q\|^2=c_1^2 n(q-1)q^{-2}.
  $$
Hence $c_1=\pm q.$ We take $K_{F_1}(x)=n(q-1)-q x_r$ choosing
$c_1=-q$ so that $K_{F_1}(0)>0.$

Next let us perform the induction step to compute $K_{F_{i+1}}(x)$:
  \begin{equation}\label{eq:gram}
   K_{F_{i+1}}(x)=c_{i+1}\Big(x_{r-i}-\sum_{j=0}^i 
               \|K_{F_j}\|^{-2}\langle x_{r-i}, K_{F_j}
    \rangle K_{F_j}(x)\Big),
  \end{equation}
where the polynomials $K_{F_j}, j=0,\dots,i,$ have the form (\ref{eq:lin})
by the induction hypothesis.
Straightforward calculations using (\ref{eq:x1})-(\ref{eq:xixi}) show that
  $$
   K_{F_{i+1}}(x)=c_{i+1}( x_{r-i}-((q-1)/q)(n-x_r-\dots-x_{r-i+1})).
  $$
Again using (\ref{eq:ortl}), we find that $c_{i+1}=\pm q^{i+1};$
as above, we will choose the minus. \qed

\medskip
(iv) The next property is a special case of (\ref{eq:flip}).
     $$
       v_e K_f(e)=v_f K_e(f)             \quad      (e,f \in \Delta_{r,n}).
     $$
In particular,
  \begin{equation}\label{eq:K0}
     K_f(0)=v_f.
  \end{equation}

\medskip
(v) For any $f,g\in \Delta_{r,n}$
    \begin{equation}\label{eq:pfgh}
      K_f(e)K_g(e)=\sum_{h\in \Delta_{r,n}} p_{f,g}^h K_h(e)
    \end{equation}
where the linearization coefficients
$p_{f,g}^h=|\{\bfz\in \cQ^{r,n}: \bfz\sim_f \bfx, \bfz\sim_g \bfy; 
\bfx\sim_h \bfy\}|$ are the intersection numbers of the scheme.
In particular, $p_{f,g}^h\geq0.$
This is a special case of property (\ref{eq:pijk}).

\medskip
(vi)  ({\em Three-term relation}) 
Let ${\mathbb K}_\kappa$ be a column vector of the polynomials $K_f$
ordered lexicographically with respect to all $f$ that satisfy $|f|=\kappa.$
The three-term relation is obtained by expanding a product 
$P(e){\mathbb K}_\kappa(e)$
in the basis $\{K_f\}$, where $P(e)$ is a first-degree polynomial. 
By orthogonality, the only nonzero terms in this expansion will be 
polynomials of degrees $\kappa+1,\kappa,\kappa-1$ \cite[p.~75]{dun01}.

We will establish an explicit form of the three-term relation for
$P(e)=\dcrit rn -|e|'.$ We have
  \begin{equation}\label{eq:3term}
    P(e) {\mathbb K}_\kappa(e)=a_\kappa {\mathbb K}_{\kappa+1}(e)
      +b_\kappa{\mathbb K}_\kappa(e)
        +c_\kappa {\mathbb K}_{\kappa-1}(e),
  \end{equation}
where $a_\kappa, b_\kappa, c_{\kappa}$ are matrices of order 
$\binom{\kappa+r-1}{r-1}\times\binom{\kappa+s+r-1} {r-1}$ and $s=1,0,-1,$ respectively.
The nonzero elements of these matrices have the following form:
   $$ 
   \begin{array}{ll}
   a_\kappa[f,h]=L_i (f_i+1)
        \quad&\text{if } h=(f_1,\dots,f_i+1,\dots,f_r),\\[2mm]
      c_{\kappa}[f,h]=L_i(n-\kappa+1)q^{i-1}(q-1) &\text{if } 
    h=(f_1,\dots,f_i-1,\dots,f_r),
\end{array}
   $$
 \begin{equation}\label{eq:mel}
b_{\kappa}[f,h]=\left\{\begin{array}{ll}
       L_i f_iq^{i-1}(q-2)
      &\text{if } h=f,\\ 
    \displaystyle L_i(f_k+1)q^{i-1}(q-1)
   &\text{if } h=(f_1,\dots,f_k+1,\dots,f_i-1,\dots,f_r),\\& \quad
         1\le k<i,\\
   \displaystyle L_i(f_i+1)q^{k-1}(q-1)
   &\text{if } h=(f_1,\dots,f_k-1,\dots,f_i+1,\dots,f_r),\\& \quad
         1\le k<i,
  \end{array}\right.
 \end{equation}    
where $L_i=\frac{q^{r-i+1}-1}{q^r(q-1)}.$ 

\medskip
\medskip{\em Proof:} According to Property (v), 
the coefficients of the expansion of the product $K_{F_i}(e)K_f(e)$ 
into the basis $\{K_h\}$ are given by the intersection numbers of the
scheme:
   \begin{equation}\label{eq:pFih}
    K_{F_i}(e)K_f(e)=\sum_h p_{F_i,f}^h K_h(e).
   \end{equation}
Because the ordered metric is
translation-invariant, we can assume that $\bfy=0,$ so $p_{F_i,f}^h$
is the number of vectors $\bfz\sim_f 0$ 
that satisfy 
$
\bfz\sim_{F_i} \bfx
$
for a fixed vector $\bfx\sim_h 0$.
In other words,
  \begin{equation}\label{eq:one}
     \bfz-\bfx=(0^r,\dots,0^r, (u_1,\dots,u_{i-1},u_i,0,\dots,0),0^r,\dots,0^r)
  \end{equation}
where the nonzero block is located in any of the $n$ possible blocks,
$u_j\in \ff_q, 1\le j<i, u_i\ne0.$

The numbers $p_{F_i,f}^h$ are nonzero only in the 
three following cases.
\begin{enumerate}
   \item $|h|=|f|+1.$  
By the above we have that $h_j=f_j$ for $j\ne i$ and $h_i=f_i+1.$
Hence $z$ can be chosen so that its $f_i$ blocks of weight $i$ 
annihilate the corresponding blocks of $x$, leaving one such block
in any of the $h_i=f_i+1$ locations.
 Thus,
     \begin{equation*}
        p_{F_i,f}^h=\begin{cases} f_i+1, &h=(f_1,\dots,f_i+1,\dots,f_r)\\
            0 &\text{otherwise}.\end{cases}
     \end{equation*}

   \medskip\item $|h|=|f|.$
 The following numbers are easily verified by (\ref{eq:one}):
   $$
    p_{F_i,f}^h = \left\{\begin{array}{ll} 
            f_i(q-2)q^{i-1}, & h=f,\\
    (f_k+1)(q-1)q^{i-1}, & 
         h = (f_1,\ldots,f_k+1,\ldots,f_i-1,\ldots,f_r), 1 \leq k < i,\\
 (f_i+1)(q-1)q^{k-1}, & h = (f_1,\ldots,f_k-1,\ldots,f_i+1,\ldots,f_r),
      1 \leq k < i,\\
  {0} & \text{otherwise.}\end{array}\right.
  $$   
 Other than these three cases, no other possibilities for $h$ arise.

  \medskip \item $|h|=|f|-1.$ Now we should add to $x$
one block of weight $i$ in any of the $n-|f|+1$ all-zero blocks. 
Thus we obtain
   $$
    p_{F_i,f}^h =(n-|f|+1)q^{i-1}(q-1) \quad h=(f_1,\dots,f_i-1,
   \dots,f_r)
   $$
and $p_{F_i,f}^h=0$ for all other $h$.
\end{enumerate}
To prove (\ref{eq:3term}) we now need to represent $P(e)$ as a linear
combination of the $K_{F_i}$s. Using (\ref{eq:lin}) we find that
   $$
     |e|'=\sum_i ie_i=\dcrit rn -\sum_{i=1}^r L_i K_{F_i}(e),
   $$
hence
   \begin{equation}\label{eq:Pe}
  P(e)=\sum_{i=1}^r L_iK_{F_i}(e).
  \end{equation}
The proof is now concluded by using (\ref{eq:pFih}) together with
the intersection numbers computed above.
\qed

Along with the polynomials $K_e$ below we use their normalized version
   $
      \tilde K_e={K_e}/{\sqrt{v_e}}.
   $
The polynomials $(\tilde K_e, e\in \Delta_{r,n})$ 
form an orthonormal basis of $V$.

Denote by $A_\kappa,B_\kappa,C_\kappa$ the coefficient matrices of
the normalized form of relation (\ref{eq:3term}). The new matrix elements
are given by
    $$
   \begin{array}{ll}
   A_\kappa[f,h]=L_i \sqrt{(f_i+1)(n-\kappa)q^{i-1}(q-1)}
        \quad&\text{if } h=(f_1,\dots,f_i+1,\dots,f_r),\\[3mm]
      C_{\kappa}[f,h]=L_i\sqrt{(n-\kappa+1)f_iq^{i-1}(q-1)} &\text{if } 
    h=(f_1,\dots,f_i-1,\dots,f_r),
\end{array}
   $$  
 \begin{equation}\label{eq:B_h}
B_{\kappa}[f,h]=\left\{\begin{array}{ll}
       L_i f_iq^{i-1}(q-2)
      &\text{if } h=f,\\[3mm]
    \displaystyle L_i\frac{q-1}q\sqrt{(f_k+1)f_iq^{k+i}}
   &\text{if } h=(f_1,\dots,f_k+1,\dots,f_i-1,\dots,f_r),\\& \quad
         1\le k<i,\\
   \displaystyle L_i\frac{q-1}q\sqrt{f_k(f_i+1)q^{k+i}}
   &\text{if } h=(f_1,\dots,f_k-1,\dots,f_i+1,\dots,f_r),\\& \quad
         1\le k<i.
  \end{array}\right.
 \end{equation}

Let $V_\kappa\subset V$ be the set of polynomials of total degree 
$\leq \kappa.$ Let $E_\kappa$ be the orthogonal projection of
$V$ on $V_\kappa$. Define the operator
    \begin{align*}
         S_\kappa: & V_\kappa\to V_\kappa\\
                   & f \mapsto E_\kappa (Pf).
   \end{align*}
Its matrix in the orthonormal basis has the form
     \begin{equation*}
      \tilde\bfS_\kappa = \left[\begin{array}{ccccc}
        B_0     & A_0   & 0          & \ldots &0\\
        C_1     & B_1   & A_1       & \ldots &0\\
        0       & C_2   & B_2      & \ldots &0\\
        \vdots  &\vdots &\vdots  & \ddots &\vdots\\
        0       & 0     &\ldots  &C_\kappa & B_\kappa\end{array}\right]
\end{equation*} 
where the $B_i$s are symmetric and $C_i=A_{i-1}^T,i=1,\dots,\kappa.$ 
On account of property (v) and (\ref{eq:Pe}), the matrix elements of 
$\tilde\bfS_\kappa$ are nonnegative. 

The matrix of $S_\kappa$ in the basis $\{K_f\}$ has the property
       \begin{equation}\label{eq:nonsymm}
          v_h \bfS_{\kappa}[f,h]=v_f \bfS_{\kappa}[h,f] 
                      \quad(f,h\in\Delta_{r,n}).
       \end{equation}

\medskip
\nd(vii) ({\em Explicit expression})
  \begin{equation}\label{eq:expl}
    K_f(x)=q^{|f|'-|f|} \prod_{i=1}^r k_{f_{i}}(n_i,x_{r-i+1}),
  \end{equation}
where $k_{f_i}$ is a univariate Krawtchouk polynomial
(\ref{eq:k}), $n_i=\sum_{j=0}^{r-i+1} x_j-\sum_{j=i+1}^r f_j,$ 
and $f,x\in \Delta_{r,n}.$ This form of the polynomial $K_f(x)$ was 
obtained in \cite{bie06a} (various other forms were found in 
\cite{mar99,dou02}). 
We remark that (\ref{eq:expl}) can be proved by performing the Gram-Schmidt 
procedure (\ref{eq:gram}) for monomials of higher degrees.
It is known that the resulting system of polynomials is unique up to a 
constant factor 
once the polynomials of degrees 0 and 1 together with the three-term relation
(\ref{eq:3term}) have been fixed, see \cite[Theorem 3.4.9]{dun01}.

 \medskip(viii)
The matrix elements of the eigenvalue matrices of the association scheme $\hr$
are given by $P_{fe}=Q_{fe}=K_f(e).$ This follows from the previous property 
and (\ref{eq:ort1}) because the polynomials $\{K_f\}$ form a unique 
orthogonal family on $\Delta_{r,n}$ with respect to the weight $w(e).$

\medskip(ix) ({\em Fourier transform representation}) Let $\omega$
be a $q$th degree primitive root of unity, $e,f\in \Delta_{r,n}.$ 
Then 
      \begin{equation}\label{eq:Fourier}
         K_f(e)=\sum_{\bfz:\bfz\sim_f 0}\omega^{\bfx\cdot\bfz}
      \end{equation}
where $\overline{\shape}(\bfx)=e.$ 
In \cite{bie06a} this relation is taken as a definition of the polynomials 
$K_f$.
Under our approach, it follows from the well-known Fourier transform 
representation
of the Krawtchouk polynomials $k_i(n,x)$ in the case $r=1$ and Theorem 
\ref{thm:MS}.

\medskip(x) ({\em Christoffel-Darboux}). 
Let $L\subset \Delta_{r,n}$ and define
        $$
          U_L(a,e)\triangleq \sum_{f\in L} v_f^{-1} K_f(a)K_f(e) 
      \quad(a,e\in\Delta_{r,n}).
        $$
Let $\bfS$ be the matrix of the operator $S:V_n \to V_{n+1}$
given by $f \mapsto Pf,$ written in the basis $\{K_f\}.$
The action of $P(e)$ on $U_L$ is described as follows:
  $$
     (P(e)-P(a))U_L(a,e)=\sum_{f\in L}v_f^{-1}\sum_{h\in \Delta_{r,n}}
          \bfS[f,h](K_h(e)K_f(a)-K_h(a)K_f(e))
  $$
  $$
   =\sum_{f\in L}v_f^{-1}\sum_{h\in \Delta_{r,n}\backslash L}
          \bfS[f,h](K_h(e)K_f(a)-K_h(a)K_f(e)),
  $$
the last equality justified by (\ref{eq:nonsymm}) as follows:
  \begin{align*}
      &\sum_{f,h\in L} v_f^{-1}\bfS[f,h](K_h(e)K_f(a)-K_h(a)K_f(e))\\
   =&\sum_{f,h} \bfS[f,h]\sqrt{\frac{v_h}{v_f}}
       (\tilde K_h(e)\tilde K_f(a)-\tilde K_f(e)\tilde K_h(a))\\=&0.
  \end{align*}
A particular case of the above is obtained when $L=\{f: |f|\leq \kappa\}$.
The kernel $U_L$, denoted in this case by $U_\kappa$, equals
$U_\kappa=\sum\limits_{s=0}^\kappa \tilde {\mathbb K}_s(e)^T
         \tilde {\mathbb K}_s(a),$ and we obtain
      \begin{equation}\label{eq:cd}
    (P(e)-P(a)) U_\kappa(a,e)
        =\tilde {\mathbb K}_{\kappa+1}(e)^TA_\kappa^T
      \tilde {\mathbb K}_\kappa(a)-\tilde {\mathbb K}_{\kappa}(e)^TA_\kappa
         \tilde {\mathbb K}_{\kappa+1}(a)
      \end{equation}
     $$
     =\sum_{f: |f|=\kappa}
              Q_f(e)\tilde K_f(a)-\tilde K_f(e)Q_f(a),
    $$
where $Q_f(e)=\sum_{|h|=\kappa+1} \tilde K_h(e)A_\kappa[f,h].$
This relation is called the Christoffel-Darboux formula.

\medskip(xi) The generating function of the polynomials $K_f$ is given by
  $$
    \sum_f K_f(e)z^f=\Big(1+(q-1)\sum_{i=1}^r  q^{i-1}z_i\Big)^{n-|e|}
    \prod_{j=1}^r \Big(1+(q-1)\sum_{k=1}^{j-1}q^{k-1}z_k-
            q^{j-1}z_j\Big)^{e_{r-j+1}}.
  $$
In particular, 
    $$
      \sum_{f\in \Delta_{r,n}} K_f(e)=q^{rn} \delta_{e,0}.
    $$
{\em Remarks:} 1. The polynomials $K_e$ were considered in 
\cite{mar99,dou02,bie06a}. However none of these papers constructed them
from their definition as eigenvalues of the $r$-Hamming scheme (to be more
precise, Martin and Stinson \cite{mar99} mention this approach but pursue 
the path suggested in Theorem \ref{thm:MS} which makes explicit calculations
difficult). Under the approach taken above, many properties
of the polynomials $K_e$ follow as special cases of the 
general combinatorial results of the previous section.

2. Other generalizations of univariate Krawtchouk polynomials were considered earlier
in \cite{tra89,roo02}. These papers study biorthogonal 
polynomials for the weight given by the multinomial probability
distribution, resulting in polynomial families different from the one considered above.

3. Property (xi) implies a MacWilliams theorem for NRT codes. It was previously
proved in \cite{mar99,dou02} using different means.
\begin{theorem}\label{thm:mw} {\rm (MacWilliams theorem in the NRT space).}
Let $C\subset \hr$ and $C^\bot\subset \hl$ be two linear codes that
satisfy $\sum_{i=1}^{nr} x_i y_i=0$ for every $\bfx\in C,\bfy\in C^\bot.$
Let $A(z_0,z)=\sum_e A_{e}\prod_{i=0}^r z_i^{e_i}$ be the weight enumerator
of $C$ and let $A^\bot(z_0,z)$ be the same for $C^\bot.$ Then
   $$
    A^\bot(z_0,z_1,\dots,z_r)=\frac 1{|C|} A(u_0,u_1,\dots,u_r)
   $$
where
  $$
  u_0=z_0+(q-1)\sum_{i=1}^r q^{i-1}z_i, \quad u_{r-j+1}=z_0+(q-1)
     \sum_{k=1}^{j-1}q^{k-1}z_k-q^{j-1}z_j, 1\le j\le r.
  $$
\end{theorem}

\section{A Linear Programming Bound on Codes and OOAs} \label{sect:lpb}
In this section we prove one of our main results, an LP
bound on the rate of codes. Let ${\sf K}=(K_f(e))_{f,e\in 
\Delta_{r,n}}$ be the eigenvalue matrix of the $r$-Hamming scheme, 
where for definiteness we are assuming the lexicographic order on the 
partitions.
Let ${\sf A}=(A_e), e\in\Delta_{r,n},$ be a vector of nonnegative
real variables, with the same ordering. Define two linear programs,
      \begin{equation*}
       \begin{array}{c@{\hskip-.5cm}l}     
         {\rm (I)}:\;\sum_{e\ne 0} A_e  \to \max\\
          \text{subject to}\\
            ({\sf A K})_e\ge 0, &\text{for all } e\\
            A_e= 0 & |e|'\le d-1,
       \end{array} \;
            \begin{array}{c@{\hskip-.5cm}l}
            {\rm(II)}:\;\sum_{e\ne 0} A_e \to \min\\
            \text{subject to}\\
             ({\sf A K})_e= 0, & |e|'\le t\\
             ({\sf A K})_e\ge 0, & |e|'> t\\A_e\ge 0,
           \end{array}
     \end{equation*}
and let LP(I) and LP(II) be their solutions.
Let $C(nr,M,d)$ be a code and $C'(t,n,r,q)$ be an OOA of size $M'.$
Then it follows from Delsarte's work \cite{del73} that 
   $$
     M\le LP{\rm (I)}, \quad M'\ge LP{\rm (II)}.
   $$
Replacing the LP problems by their duals, we obtain the following theorem.
\begin{theorem}\label{thm:lp} Let $F(x)=F_0+\sum_{e\ne 0} F_eK_e(x)$ 
be a polynomial that satisfies 
  \begin{equation}\label{eq:limits}
     F_0>0, \quad F_e\ge 0\;\; (e\ne 0); \quad F(e)\le 0 \;\;\text{ for all $e$ such that }
       |e|'\ge d.
  \end{equation}
Then any $(nr,M,d)$ code satisfies
   \begin{equation}\label{eq:codebound}
     M\le F(0)/F_0.
   \end{equation}
Any OOA of strength $t=d-1$ and size $M'$ satisfies 
  \begin{equation}\label{eq:arraybounds}
    M'\ge q^{nr} F_0/F(0).
  \end{equation}
\end{theorem}
This result is essentially due to P.~Delsarte. However, for the NRT space
it was first stated by Martin \cite{mar99,mar00b} 
and later rederived by Bierbrauer \cite{bie06a}. The fact that the 
same polynomial gives a bound both on codes and orthogonal arrays is an 
easy consequence of Delsarte's theory, first mentioned in
Levenshtein's work \cite{lev95a}.

\subsection{The bound}
In this section we use Theorem \ref{thm:lp} to derive a bound on 
ordered codes and arrays. Its proof uses a ``spectral method'' first 
employed in \cite{bac06} for the Grassmannian
space and later  used in \cite{bar06a} to prove classical asymptotic 
bounds of coding theory.
The gist of the method can be explained as follows. The polynomial 
$F(e)$ is sought in the form $F(e)=u(e)G^2(e)$ where $u(e)$ is a linear
function that assures that $F(e)\le 0$ in (\ref{eq:limits}) and $G(e)$ is 
a function that maximizes the Fourier transform $\widehat F(0).$ 
In the univariate case
it turns out that a good choice for $G$ is a delta-function at (or near)
$d$.
An approximation of the delta-function is given by the (Dirichlet) kernel
$U_\kappa$ 
which is its projection on $V_{\kappa}$. We therefore seek to modify
the operator $S_\kappa$ so that $U_\kappa$ becomes its eigenfunction with
eigenvalue $\theta_\kappa$, express the bound of Theorem \ref{thm:lp}
as a function of $\theta_\kappa$ and optimize on $\kappa$ within the
limits (\ref{eq:limits}). 
The reader is referred to \cite{bar08a} for a more detailed discussion
of these ideas.

\begin{theorem} \label{thm:bound} Let $\kappa$ be any degree such that 
$P(e)\le \lambda_{\kappa-1}$ for all shapes $e$ with $|e|'\ge d,$ where
$\lambda_i$ is the maximum eigenvalue of $S_i$ and $d\ge 1$ is an integer. 

Let $C\subset \hr$ be an $(nr,M,d)$ code. Then 
   \begin{equation}\label{eq:bound-r}
     M\le \frac {4r\dcrit(n-\kappa)(q^r-1)^\kappa}
         {\dcrit rn-\lambda_\kappa}\binom n\kappa.
   \end{equation} 
Let $C$ be a $(t=d-1,n,r,q)$ OOA of size $M$. Then
  \begin{equation}\label{eq:bound-ooa}
    M\ge \frac{q^{nr}}{\binom n\kappa}
       \frac{(\dcrit rn-\lambda_\kappa)}{4r\dcrit
      (n-\kappa)(q^r-1)^\kappa}.
  \end{equation}
\end{theorem}
\begin{proof}  Consider the operator $T_\kappa$ that equals $S_\kappa$
on $V_{\kappa-1}$ and acts 
on a function $\varphi\in V_\kappa\backslash V_{\kappa-1}$ by
   $$
    T_\kappa(\varphi):=S_\kappa \varphi-
        \sum_{f: |f|=\kappa} \epsilon_f \varphi_f \tilde K_f,
  $$
where $\epsilon_f>0$ are some constants indexed by the partitions of
weight $\kappa$ (their values will be chosen later).
The matrix of $T_\kappa$ in the orthonormal basis equals
   $$ 
       \tilde \bfT_\kappa=\tilde \bfS_\kappa - 
       \left[\begin{array}{cc}0 & 0\\ 0 & E\end{array}\right]
   $$
where $E=\text{diag}(\epsilon_f, |f|=\kappa)$ is a matrix of order $\binom{\kappa+r-1}{r-1}.$ 
 Let $m$ be such that $\tilde\bfT_{\kappa}+mI>0.$ By Perron-Frobenius \cite[p.~80]{bro89},
the spectral radius $\rho(T_\kappa+mI)$ is well defined and
is an eigenvalue of (algebraic and geometric) multiplicity one of 
$T_\kappa+mI$. Moreover, again using Perron-Frobenius,
   $$
      \rho(S_{\kappa-1}+mI)<\rho(T_\kappa+mI)<\rho (S_{\kappa}+mI).
   $$
Then
   \begin{equation}\label{eq:lambda}
     \lambda_{\kappa-1}<\theta_\kappa<\lambda_\kappa
   \end{equation}
where $\theta_\kappa=\rho(T_\kappa).$ Let $G>0$ be the eigenfunction of 
$T_\kappa$ with eigenvalue $\theta_\kappa.$ 
Let us write out the product $P(e)G$ in the orthonormal basis:
   $$
     P(e)G=
       S_\kappa G+
        G_\kappa A_\kappa \tilde {\mathbb K}_{\kappa+1}
        =\theta_\kappa G+\sum_{f: |f|=\kappa} \epsilon_f 
               G_f \tilde K_f
       +G_\kappa A_\kappa \tilde {\mathbb K}_{\kappa+1}.
   $$
where $G_\kappa$ is a projection of the vector $G$ on the space 
$V_\kappa\backslash V_{\kappa-1}.$ 
This implies the equality
   $$
     G=\frac{\sum_{|f|=\kappa} G_f(\epsilon_f \tilde K_f+Q_f)}
          {P(e)-\theta_\kappa},
   $$
where $Q_f(x)$ is defined after (\ref{eq:cd}). Now take 
$F(x)=(P(x)-\theta_\kappa)G^2(x).$ Let us verify (\ref{eq:limits}).
Since multiplication by a function is a self-adjoint operator, we obtain
  $$
    F_0=\langle F,1\rangle=\langle \sum_{|f|=\kappa}G_f(\epsilon_f
      \tilde K_f+Q_f),G\rangle = \sum _{|f|=\kappa} G_f^2\epsilon_f>0.   
  $$
Using (\ref{eq:pfgh}) one can easily check that $F_e\ge 0$ for all $e\ne 0.$
The assumption of the theorem together with (\ref{eq:lambda}) implies
that $F(f)\leq 0$ for $|f|'\geq d.$ Hence
      \begin{align*}
      M&\leq \frac{F(0)}{F_0}\nonumber \\
       &= \frac{\Big(\sum_{|f|=\kappa} G_f(\epsilon_f \tilde K_f(0)
                       +Q_f(0))\Big)^2}
           {(P(0)-\theta_\kappa)\sum_{|f|=\kappa} G_f^2\epsilon_f \nonumber }\\
       &\leq \frac 1{P(0)-\lambda_\kappa}\sum_{|f|=\kappa}\frac{
       (\epsilon_f \tilde K_f(0)+Q_f(0))^2}{\epsilon_f},
     \end{align*}
where in the last step we used the Cauchy-Schwarz inequality.
Computing the minimum on $\epsilon_f,$ we obtain
    \begin{equation}\label{eq:F0}
       M \leq \frac 4{P(0)-\lambda_\kappa}\sum_{|f|=\kappa} Q_f(0)
         \sqrt{v_f}.
   \end{equation}
Next, 
   $$\sum_{|f|=\kappa} Q_f(0)\sqrt{v_f}=
      \sum_{f: |f|=\kappa} \sqrt{v_f}
     \sum_{h:|h|=\kappa+1} A_\kappa[f,h]\sqrt{v_h}.
   $$     
Let $h=(f_1,\dots,f_i+1,\dots,f_r)$ for some $i, 1\le i\le r.$ Then
using (\ref{eq:v}) we find
   \begin{align*}
     A_\kappa[f,h] \sqrt{v_h}&=
            L_i\sqrt{(f_i+1)(n-\kappa)q^{i-1}(q-1)}\sqrt{v_h}\\
    &=L_i\sqrt{(f_i+1)(n-\kappa)q^{i-1}(q-1)}
         \sqrt{v_f\frac{(n-\kappa)q^{i-1}(q-1)}{f_i+1}}\\
     &=\Big(1-\frac1 {q^{r-i+1}}\Big)(n-\kappa)\sqrt{v_f}.
   \end{align*}
Thus we have 
   $$
    \sum_{|f|=\kappa} Q_f(0)\sqrt{v_f}=\sum_{|f|=\kappa}
  \sum_{i=1}^r(n-\kappa)\Big(1-\frac1 {q^{r-i+1}}\Big) v_f
   $$
   $$
       =(n-\kappa)r\dcrit\sum_{|f|=\kappa}v_f=(n-\kappa)r\dcrit
             \binom n\kappa(q^r-1)^\kappa.
   $$
Substitution of this expression into (\ref{eq:F0}) concludes the proof of
(\ref{eq:bound-r}). The bound (\ref{eq:bound-ooa}) follows by 
(\ref{eq:arraybounds}).
\end{proof}   

\subsection{Spectral radius of $\bfS_{\kappa}$}
In this section we derive an asymptotic lower bound on 
the spectral radius of $\bfS_\kappa.$ 
This estimate will be later used to optimize the bound (\ref{eq:bound-r}) 
on the choice of $\kappa.$

\begin{theorem}\label{lemma:lambda}
     $$
     \lim_{\substack {n\to\infty\\ \frac{\kappa}{n}\to\tau}}
       \frac{\lambda_\kappa}{n} \ge  \max_{\substack {\tau_i \geq 0\\
\sum_{i=1}^r\tau_i = \tau}} \Lambda(\tau_1,\dots,\tau_r), 
   $$ 
where
     \begin{align}
    \Lambda(\tau_1,\dots,\tau_r)=\sum_{i=1}^r L_i
&\Big(2\sqrt{(1-\tau)\tau_i(q-1)q^{i-1}}\nonumber\\[-2mm]    
      &+  (q-2)\tau_i (q^{r}-q^{i-1}) +2\frac{(q-1)}{q}
\sum_{k=1}^{i-1}\sqrt{\tau_k\tau_i q^{i+k}}\Big).\label{eq:G}
     \end{align}
 \end{theorem} 
To prove this theorem, we will bound below the largest eigenvalue 
$\lambda_{\kappa}$ of the matrix $\tilde \bfS_{\kappa}.$ 
 For any real vector $y$ we have
   $$
    \lambda_{\kappa} \ge\frac{y^T \tilde \bfS_{\kappa} y}{(y,y)}.
   $$
We will construct a suitable $(0,1)$-vector $y.$
Its coordinates are indexed by the partitions arranged in the increasing order
of their length $\mu$ and lexicographically within a block of coordinates
for each value of $\mu, 0\le \mu\le \kappa$. Let $y=(y_0,y_1,\dots,y_\kappa)$
where $y_\mu=(y_f, |f|=\mu).$

Let $f,|f|=\mu,$ be a shape vector. For an integer $J$ consider the set
   $$
    \cF_{\mu}=\cF_\mu(J,f)\triangleq \{(f_1+ \mu - \kappa +j_1,\dots,f_r+j_r):  
     \sum_{i=1}^r j_i=0;\; |j_i|\le J, i=1,\dots,r\}
   $$
and denote $m=|\cF_\mu|.$ Next, let
    $$(y_\mu)_h=1(h\in \cF_\mu)$$
for $\mu=\kappa+1-s,\dots,\kappa$ where $s$ will be chosen later, and
$y_\mu=0$ otherwise.

In the next two lemmas we derive a lower bound on the part
of the product $y^T \tilde \bfS_{\kappa}y$ that involves only the rows
of $\tilde \bfS_{\kappa}$ that correspond to the shapes $f$ of length $\mu.$
 Let
   $$
     E_h=\{(h_1,\dots,h_k\pm 1,\dots,h_l\mp1, \dots,h_r),
       1\le k<l\le r\}
  $$
be the index set of the nonzero off-diagonal elements in the row in $B_\mu$ which is indexed by $h=(h_1,\ldots,h_r)$.
\begin{lemma} Let $e=\text{\rm argmin}_{h\in \cF_{\mu}}
(\sum_{g\in E_h\cup\{h\}}   B_{\mu}[h,g])$ and let 
  $\psi_\mu=\sum_g B_{\mu}[e,g].$ Then
         $$
  y_{\mu}^T B_\mu y_\mu \ge \psi_\mu m(1-o_m(1)) .
      $$
\end{lemma}
\begin{proof} 
I. Since $|h|=\mu$ for every $h\in \cF_{\mu},$ the quantity
$|\cF_{\mu}|$ equals the number of ordered partitions of 0 into at most
$r$ parts, each part bounded between $-J$ and $J$, or the number of 
ordered partitions
   $$
     Jr=\sum_{i=1}^r j_i,\quad 0\le j_i\le 2J, \;i=1,\dots,r.
   $$
The number of such partitions is given by
\cite[p.~1037]{ges95}:
  $$
    \pi(r,2J,Jr)=\sum_{i=0}^{\lfloor \frac r2\rfloor}
    (-1)^i\binom ri\binom{r+Jr-(2J+1)i-1}{r-1}.
  $$
Writing this expression as a polynomial in $J$, we find the coefficient
of $J^{r-1}$ to be
  $$
    \frac 1{(r-1)!}\sum_{i=0}^{\lfloor \frac r2\rfloor} (-1)^i 
\binom ri (r-2i)^{r-1}.
  $$
Since this is always positive\footnote{
To prove positivity, observe that the numbers 
$S_{r,m}=\sum_{i=0}^{\lfloor \frac r2\rfloor} (-1)^i \binom ri (r-2i)^{m}$
satisfy the recurrence
   $$
  S_{r,m}=r^2 S_{r,m-2}+4r(r-1)S_{r-2,m-2}, \quad 3\le m\le r-1
  $$
and then use induction to prove that $S_{r.m}>0$ $(<0)$ according as
$r-m\equiv 1$ or 3 mod 4.
}, we conclude that
$m$ is a degree-$(r-1)$ polynomial in $J$; in particular, if $J\to\infty,$
then also $m\to\infty$.

II. 
We have
   $$
    y_\mu^T B_\mu y_\mu =\sum_{h,g\in \cF_\mu} B_\mu[h,g].
   $$
To bound $y_\mu^T B_\mu y_\mu$ below we estimate the difference between
the above sum and the sum of all the nonzero elements of $B_\mu$ in 
the rows $h\in \cF_\mu$ which is obtained by replacing the range of
column indices $g$ above with $g\in E_h\cup\{h\}.$ Therefore, for a given
$h\in \cF_\mu$ let us
estimate the number
$|E_h\backslash \cF_{\mu}|$ of nonzero entries in $B_\mu[h,\cdot]$ not 
included in the sum. Let $f=(f_1,\dots,f_r)$ and let $h$ be of the form
    $
        h=(\dots,f_k+J,\dots)\in \cF_\mu
    $
for some $1\le k\le r$.
Consider the column indices $g\in E_h$ given by
   \begin{equation}\label{eq:+}
    g=(f_1+\mu-\kappa+j_1,\dots,f_k+J+1,\dots,f_l+j_l-1,\dots,f_r+j_r)
  \end{equation}
for any $k\ne l\in \{1,\ldots,r\}.$ For any pair $h,g$ of this form, $B_\mu[h,g]\ne 0$
but $g\not\in \cF_\mu.$
The number of shapes $h$ that result in shapes $g$ of the form (\ref{eq:+}) equals 
the number of ordered partitions
of $-J$ into at most $r-1$ parts of magnitude $\leq J$; equivalently, this 
is the number of ordered partitions 
     $$
         J(r-2)=j_2+\dots+j_r,\quad 0\le j_i\le 2J, i=2,\dots,r,
     $$
which equals $\Pi_+\triangleq\pi(r-1,2J,J(r-2)).$

Next consider the row indices $h=(\dots,f_k-J,\dots)\in \cF_\mu$ and column 
indices $g\in E_h$ given by
    \begin{equation*}\label{eq:-}
    g=(f_1+\mu-\kappa+j_1,\dots,f_k-J-1,\dots,f_l+j_l+1,\dots,f_r+j_r)
  \end{equation*}
which again account for $B_\mu[h,g]\ne 0$ and $g\not\in \cF_\mu.$
The number of such shapes $h$ equals the number of ordered partitions
of $Jr$ into $r-1$ or fewer parts $0\le j_i\le 2J.$ Denote this number
by $\Pi_-\triangleq\pi(r-1,2J,Jr).$ Note that as $J\to\infty,$
both $\Pi_+$ and $\Pi_-$ grow proportionally to $J^{r-2}.$

It is easy to verify that $E_h\backslash \cF_{\mu}\ne\emptyset$ if and 
only if $h$ and $g$ are of the described form. Observe that
by (\ref{eq:B_h}), $|E_f|=r^2-r.$ We then obtain
   \begin{align*} 
   \sum_{h,g\in \cF_\mu} B_\mu[h,g]\ge \psi_\mu(m-r(r^2-r)(\Pi_++\Pi_-))
    =  \psi_\mu m(1-o_m(1)).
   \end{align*}
The lemma is proved.\hfill\end{proof}

We now consider the part of the product $y^T\tilde \bfS_\kappa y$ that
involves the matrix $C_\mu, \mu=\kappa-s+2,\dots,\kappa.$ 
 For a shape $h$ let 
  $$
   D_h=\{(h_1,\dots,h_k-1,\dots,h_r), 1\le k\le r\}.
  $$
The proof of the next lemma is very similar to the above proof and
will therefore be omitted.
\begin{lemma} Let $e=\text{\rm argmin}_{h\in \cF_{\mu}}
(\sum_{g\in D_h}   C_{\mu}[h,g])$ and let 
  $\phi_\mu=\sum_g C_{\mu}[e,g].$
    $$
      y_\mu^T C_\mu y_{\mu-1}\ge \phi_\mu m(1-o_m(1)).
   $$
\end{lemma}

To complete the proof of Theorem \ref{lemma:lambda}, compute
   \begin{align*}
     \lambda_\kappa&\ge \frac1{ms} y^TS_\kappa y\\
  &=\frac 1{ms}\Big(\sum_{\mu=\kappa+1-s}^{\kappa} 
       y_\mu^T B_\mu y_\mu
       +2\sum_{\mu=\kappa+2-s}^{\kappa} y_\mu^T C_\mu y_{\mu-1}\Big)\\
  &\ge \frac1s\Big(\sum_{\mu=\kappa+1-s}^{\kappa} \psi_\mu+
      2\sum_{\mu=\kappa+2-s}^{\kappa} \phi_\mu\Big)(1-o_m(1))\\
  &\ge  \psi^\ast+2\frac{s-1}s\phi^\ast (1-o_m(1)),
   \end{align*}
where $\psi^\ast$ ($\phi^\ast$) is the smallest of the numbers $\psi_\mu$
($\phi_\mu$) above. Note that both $\psi^\ast$ and $\phi^\ast$ are nonzero.
Now let $n\to\infty,\kappa=\tau n,$ and let us choose $f$ in the definition of 
$\cF_\mu$ to be of the form $f=(f_1,\dots,f_r), f_i=n\tau_i, 1\le i\le r.$
We assume that none of the $\tau_i$'s approach $0$ as $n$ grows.
Take $s=o(n), s\to\infty.$ Using (\ref{eq:B_h}), and letting $J=o(n),
J\to\infty$ we get
   $$
  \lim_{n\to\infty} \frac{\psi^\ast} n =\sum_{i=1}^r L_i
    \Big((q-2)\tau_i (q^{r}-q^{i-1}) +2\frac{(q-1)}{q}
          \sum_{k=1}^{i-1}\sqrt{\tau_k\tau_i q^{i+k}}\Big)
  $$
  $$
     \lim_{n\to\infty} \frac{\phi^\ast} n =\sum_{i=1}^r L_i
          \sqrt{(1-\tau)\tau_i(q-1)q^{i-1}}.
  $$
Then since $\kappa/n \to\tau,$
  $$
  \lim_{n\to\infty} \frac {\lambda_\kappa}n\ge \lim_{n\to\infty}
      \frac{y^T S_\kappa y} {msn} \ge \Lambda(\tau_1,\dots,\tau_r).
  $$
The theorem is proved.

\subsection{Asymptotic estimate for codes and OOAs}

Theorems \ref{thm:bound} and \ref{lemma:lambda} together enable us
to prove one of the main results of the paper.
\begin{theorem} Let $R_{\text{\rm LP}}(\delta)$ be the function
defined parametrically by the relations
   \begin{align}
     R(\tau)&= \frac1r \left(
                h_q(\tau)+\tau\log_q\frac{q^r-1}{q-1}\label{eq:boundR}
                \right)\\
    \delta(\tau)&=\dcrit-\frac 1r \max\limits_{\substack {\tau_i \geq 0\\
\sum_{i=1}^r\tau_i = \tau}} \Lambda(\tau_1,\dots,\tau_r),
               \quad 0\le\tau\le 1. \label{eq:boundd}
    \end{align} 
Then the asymptotic rate of any code family
of relative distance $\delta$ satisfies
$R\le R_{\text{\rm LP}}(\delta)$ and the rate of any family of OOAs of relative
strength $\delta$ satisfies $R\ge 1-R_{\text{\rm LP}}(\delta).$
\end{theorem}
To prove this theorem, take the logarithms in (\ref{eq:bound-r})
and pass to the limit as $n\to\infty.$ Using the standard asymptotics for
the binomial coefficient, we find that the code rate is bounded above by
the right-hand side of (\ref{eq:boundR}). 
The condition $P(e)\le \lambda_{\kappa-1}$ 
of the Theorem \ref{thm:bound} will be satisfied for large $n$ if
  $$
     \dcrit-\delta\le \frac {\lambda_{\tau n}}{rn}.
  $$
This defines the function in (\ref{eq:boundd}). Thus, the proof is complete.

{\em Remark.} For $r=1$ this bound reduces to the linear programming
bound on the rate of codes in \cite{aal77}. 
Just as that result, the bound of this theorem improves upon the asymptotic
Plotkin bound for large values of the code distance.

\section{The Case $r=2$}
In this section we prove a bound for codes in $\cQ^{2,n}$ which improves
upon the general result of the previous section.
The improvement is due to the fact that in the
case $r=2$ it is possible to work with the polynomials $K_f(e)$ in their
explicit form, and base the bound on the behavior of their zeros instead
of the spectral radius of the operator $S_{\kappa}.$
Namely, let $f=(f_1,f_2),e=(e_1,e_2).$ From (\ref{eq:expl}) we have
   $$
    K_f(e)=q^{f_2} k_{_{f_2}}(n-e_2,e_1)k_{_{f_1}}(n-f_2,e_2).
   $$
We also have
  $$
    P(e)= n\Big(2-\frac{q+1}{q^2}\Big)-e_1-2e_2.
   $$
We will use the following properties of the polynomials $k_s$ whose proofs 
are found for instance in \cite{lev95a}. Let $x_i(n,s), i=1,\dots,s$
be the roots of $k_s$ in the ascending order. Then
  \begin{align} 
    &x_i (n-1, s) < x_i (n,s) < x_i (n-1, s-1) < x_i (n,s-1) < x_{i+1} (n,s),
        \label{eq:roots}\\
    &\qquad\qquad\qquad\qquad 1<s<n,\ i=1,\ldots,s-1 \nonumber.
\end{align}
Let $n\to\infty, s/n\to y.$ Then
     \begin{align}\label{eq:limrt}
     \lim_{n\to\infty} \frac{x_1(n,s)} n=\gamma(y)
     \triangleq \frac{q-1}{q} - \frac{q-2}{q}y - 
              \frac{2}{q}\sqrt{(q-1)y(1-y)}.
  \end{align}
The Krawtchouk polynomials satisfy the recurrence
   \begin{equation}
     k_s(n, x) = k_s (n -1,  x) + (q-1)k_{s-1}(n-1,x) \label{eq:k3term}
   \end{equation}    
and a Christoffel-Darboux formula of the form (\ref{eq:cd})
   \begin{equation}\label{eq:cdk}
     q(x-y)\sum_{s=0}^h \frac{k_s(x)k_s(y)}{k_s(0)}=
          \frac{h+1}{k_h(0)}(k_{h+1}(y)k_h(x)-k_{h+1}(x)k_h(y)).
        \end{equation}

{\em Remark:} Properties (\ref{eq:roots})-(\ref{eq:cdk}) are usually
stated for integer $n$. This is related to the fact that the polynomials
$k_s(n,x)$ represent the eigenvalues of the Hamming association scheme.
As pointed out to us by M. Aaltonen \cite{aal07}, it is possible to prove these
properties for any $n\in \reals^+$ relying on the generating function
of the Krawtchouk polynomials. 

The main result of this section is given in the following theorem.
\begin{theorem}\label{thm:2} 
The asymptotic rate of any family of codes of relative distance $\delta$
satisfies $R\leq \Phi(\delta)$, where
  $$
  \Phi(\delta) =\min_{\tau_1,\tau_2} \half
  \big\{\tau_2+ h_q (\tau_1) + 
(1-\tau_1) h_q \big(\frac{\tau_2}{1 - \tau_1})\big\},
  $$
where the minimum is taken over all $\tau_1,\tau_2$ that satisfy
 \begin{align*}
      &0\le \tau_1 \le (q-1)/{q^2}, \quad
        0\le \tau_2 \le (q-1)/q,\\
 &\gamma(\tau_2)+(2-\gamma(\tau_2))(1-\tau_2)\gamma(\tau_1) \le 2\delta .
  \end{align*}
The asymptotic rate of any family of OOAs of relative strength $\delta$ 
satisfies $R\ge 1-\Phi(\delta).$
\end{theorem}
The remainder of the section is devoted to the proof of this result.
We note that the polynomials $K_f(e)$ are formed as products of two 
Krawtchouk polynomials. A similar situation arose in \cite{aal90} 
which dealt with the Johnson association scheme whose (second) eigenvalues
are equal to a product of a Krawtchouk and a Hahn polynomial. 
Therefore, we adopt some elements of the analysis in \cite{aal90} in our
proof below.

In quest of an LP bound, we require a polynomial $F(e)=
 F(e_1,e_2)$ that satisfies conditions (\ref{eq:limits}). Consider
the polynomial of the form
   \begin{equation}\label{eq:F}
     F(e)=(P(e)-P(a)) (U_L(a,e))^2
   \end{equation}
for some $a=(\alpha,\beta)$ and a subset $L$. For brevity below
we write $S_{fh}$ instead of $\bfS_\kappa[f,h]$ and denote 
$\bar L=\Delta_{2,n}\backslash L.$
We find
    \begin{align*}
   F_0&=\langle F,1\rangle=\langle (P(e)-P(a))U_L(a,e),U_L(a,e)\rangle\\
    &=\langle \sum_{f\in L} \frac 1{v_f}\sum_{h\in \bar L}
       S_{fh}(-K_h(a))K_f(e), U_L\rangle\\
   &=-\sum_{f\in L}\sum_{h\in\bar L}
     S_{fh}K_h(a) \frac {K_f(a)}{v_f}.
    \end{align*}
In order to ensure that $F_0>0$ we will choose $L$ and $a$ so that
   \begin{equation}\label{eq:cond}
     K_h(a)\le 0 \quad\text{if }h\in \bar L;
   \qquad K_f(a)>0 \quad\text{if } f\in L.
   \end{equation}
Let $s=(s_1-1,s_2)\in \Delta_{2,n}$ be a shape that satisfies
$\{(s_1-1,s_2+1),(s_1,s_2)\}\subset \Delta_{2,n}.$ Let $a=(\alpha,\beta)$
satisfy
    \begin{equation}\label{eq:a}
       \beta=x_1(n-s_2,s_1),\;\quad
             x_1(n-\beta,s_2+1)<\alpha< x_1(n-\beta,s_2),\quad \alpha+2\beta
              \le d.
    \end{equation}
 For any $f_2, 0\le f_2\le s_2+1$ denote by $\phi(f_2)$ the degree such that
   \begin{equation}\label{eq:beta}
     x_1(n-f_2,\phi(f_2)+1)\le \beta< x_1(n-f_2,\phi(f_2)).
   \end{equation}
By (\ref{eq:roots}), $\phi(\cdot)$ is well defined and implies the 
following:
  $$ 
    [(x_1(n-u,w)>\beta) \; \Rightarrow \; (w\le\phi(u))], \quad
           [(x_1(n-u,w)\le\beta) \; \Rightarrow \;(w\ge \phi(u)+1)].
  $$ 
We choose the region $L$ to be given by
   $$
     L=(f_1,f_2: f_2=0,\dots,s_2; f_1=0,\dots,\phi(f_2)).
   $$
 For the moment this choice is not unique because there are many possibilities
for $s$. This ambiguity will be later removed by optimizing the bound on the
choice of $s$. 

To argue about the sign of $F_0$ we need to establish some properties
of the region $L$. First, we claim that for a fixed $f_2$,
   \begin{equation}\label{eq:ff}
       \phi(f_2)-1\le \phi(f_2+1)\le \phi(f_2).
   \end{equation}
Indeed, by (\ref{eq:roots}), 
   $$
     \beta< x_1(n-f_2,\phi(f_2))<x_1(n-f_2-1,\phi(f_2)-1)
   $$
which implies the left-hand side of (\ref{eq:ff}). On the other hand,
  $$
    \beta\ge  x_1(n-f_2,\phi(f_2)+1)> x_1(n-f_2-1,\phi(f_2)+1)
  $$
which implies the right-hand side.

\begin{figure}[ht]
       \centering
       \scalebox{0.48}{\includegraphics[keepaspectratio]{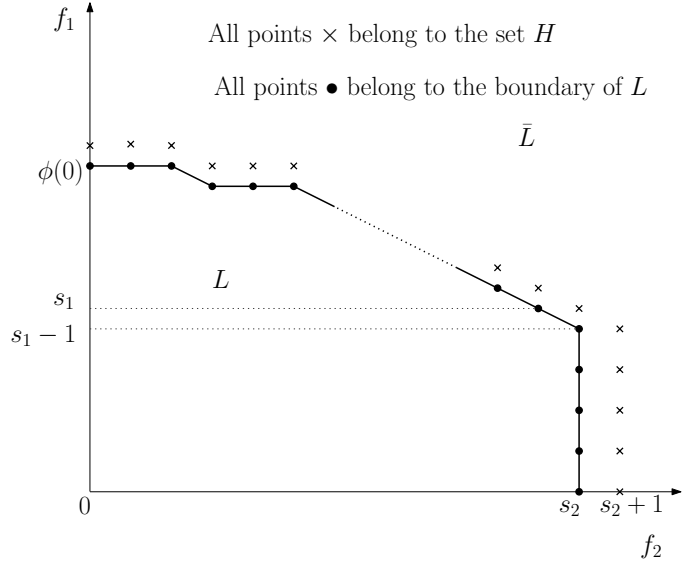}}
       \caption{The region $L$}
       \label{fig:region_L}
\end{figure}

The values of $f,h$ for which $S_{fh}\ne 0$ are given in
(\ref{eq:B_h}). In particular, if $f\in L$, then the set $H$ of the
shape vectors $h$ that index the nonzero matrix elements of $\bfS$ and that lie outside the region $L$ is as
follows:
  $$
    H=\{(\phi(f_2)+1,f_2), f_2=0,1,\dots,s_2\}\cup
         \{(f_1,s_2+1), f_1=0,1,\dots, s_1-1\}.
  $$
The region $L$ and the corresponding set $H$ are shown in 
 Fig.~\ref{fig:region_L}. By our choice of the parameters,
   \begin{align*}
    k_{f_2}(n-\beta,\alpha)&> 0 \qquad(0\le f_2\le s_2),\\ 
               k_{s_2+1}(n-\beta,\alpha)&<0,\\
     k_{f_1}(n-f_2,\beta)&>0 \qquad
             (0\le f_1\le \phi(f_2), \,0\le f_2\le s_2+1),\\
       k_{\phi(f_2)+1}(n-f_2,\beta)&\le 0 \qquad (0\le f_2\le s_2+1).
 \end{align*}

Then
  \begin{equation}\label{eq:sgn1}
    K_{(f_1,f_2)}(a)=q^{f_2} k_{f_2}(n-\beta,\alpha)k_{f_1}(n-f_2,\beta)>0,
          \;\; f\in L,
  \end{equation}
  \begin{equation}\label{eq:sgn2}
    K_{(\phi(f_2)+1,f_2)}(a) =q^{f_2}k_{f_2}(n-\beta,\alpha)k_{\phi(f_2)+1}
           (n-f_2,\beta)\le 0, \;\;0\le f_2\le s_2,
  \end{equation}
  \begin{equation}\label{eq:sgn3}
K_{(f_1,s_2+1)}(a) =q^{s_2+1}k_{s_2+1}(n-\beta,\alpha)k_{f_1}(n-s_2-1,\beta)<0,
   \;\;0\le f_1\le s_1-1.
  \end{equation}
Thus,
  \begin{equation}\label{eq:sgn}
   K_f(a)\le 0 \;\;(f\in\bar L), \quad
       K_f(a)>0 \;\;(f\in L).
  \end{equation}
This proves that $F_0>0.$

Let us show that $F_e\ge0$ for all $e.$ For this rewrite $F$ as follows:
  \begin{align*}
       &F(e)=((P(e)-P(a)) U_L(a,e)^2\\
   &=\sum_{f\in L}\frac 1{v_f}\sum_{h\in \bar L}
       S_{fh}(K_h(e)K_f(a)-K_h(a)K_f(e))\sum_{g\in L}\frac{K_g(a)
         K_g(e)}{v_g}\\
   &=\Big(\sum_{h\in \bar L}K_h(e)
   \sum_{f\in L}\frac{S_{fh}  K_f(a)}{v_f}-\sum_{f\in L}K_f(e)
      \sum_{h\in \bar L}\frac{K_h(a)S_{fh}}{v_f}
            \Big)\sum_{g\in L}\frac{K_g(a)K_g(e)}{v_g}\\
   &=\sum_{h\in \bar L,\;g\in L}\frac{K_g(a)}{v_g}
       K_h(e)K_g(e)\sum_{f\in L}\frac{S_{fh}  K_f(a)}{v_f}\\
        &\hspace*{2cm}-\sum_{f,g\in L} \frac{K_g(a)}{v_g}K_f(e)K_g(e)
         \sum_{h\in \bar L}
          \frac{K_h(a)S_{fh}}{v_f}.
   \end{align*}
By (\ref{eq:pfgh}), the products $K_hK_g$ and $K_fK_g$ 
are expanded in the basis $\{K_f\}$ with nonnegative coefficients.
Moreover, the other terms in the above formula also have the needed signs
on account of (\ref{eq:sgn1})-(\ref{eq:sgn3}). This establishes our claim.

 Finally, because of the third condition in (\ref{eq:a}), $F(e)\le 0$ for
all $e$ with $|e|'\ge d.$ 

We are now able to formulate the bound on codes and OOAs.
\begin{theorem} \label{thm:fb2} Let $C$ be an  $(2n,M,d)$ code 
$C\subset \hr(q,n,2).$ Then
\begin{equation}\label{eq:fb2}
    M\le \frac{4(n-\beta-s_2)(n-s_2-s_1+1)^2(q-1)^3 (\alpha+ 2\beta)}{q^3\alpha^2\beta^2}v_s,
\end{equation}
where 
$s=(s_1-1,s_2)$ satisfies
$\{(s_1-1,s_2+1),(s_1,s_2)\}\subset \Delta_{2,n}$ and 
$a=(\alpha,\beta)$ is chosen to fulfill conditions (\ref{eq:a}).

Let $C$ be a $(t=d-1,n,2,q)$ OOA of size $M$. Then
  \begin{equation}\label{eq:ab2}
    M\ge \frac{q^{nr}}{v_s} \frac{q^3\alpha^2\beta^2 } 
        {4(n-\beta-s_2)(n-s_2-s_1+1)^2(q-1)^3 (\alpha + 2\beta)}.
  \end{equation}
\end{theorem}
\begin{proof}
Let us compute $F_0=\langle F,1\rangle.$ Denote $\sigma_1=(s_1-1,s_2+1),
\sigma_2=(s_1-2,s_2+1).$ By (\ref{eq:sgn}) and (\ref{eq:mel}) we have
  \begin{align}
    F_0=&-\sum_{f\in L}\frac {K_f(a)}{v_f}\sum_{h\in\bar L}
     S_{fh}K_h(a)  \nonumber \\
      &\ge - \frac {K_s(a)}{v_s}
          (S_{s,\sigma_1}K_{\sigma_1}(a)+S_{s,\sigma_2}K_{\sigma_2}(a)
    +S_{s,(s_1,s_2)} K_{(s_1,s_2)}(a))\nonumber\\
  &=- \frac {K_s(a)}{q^2v_s}\Big((s_2+1)K_{\sigma_1}(a)+
        (s_2+1)(q-1)
        K_{\sigma_2}(a)+ s_1(q+1)K_{(s_1,s_2)}(a)\Big)\nonumber\\
   &=- \frac {K_s(a)(s_2+1)q^{s_2+1}}{q^2v_s}
          k_{s_2+1}(n-\beta,\alpha)k_{s_1-1}(n-s_2,\beta).\label{eq:f0}
         \end{align}
Let us now evaluate $U_L(a,0)$.
   $$
     U_L(a,0)=\sum_{f\in L}K_f(a)=\sum_{f_2=0}^{s_2}q^{f_2}k_{f_2}
   (n-\beta,\alpha) \sum_{f_1=0}^{\phi(f_2)}k_{f_1}(n-f_2,\beta).
   $$
Let us bound above the last sum. We shall prove that 
   \begin{equation}\label{eq:in}
   \sum_{f_1=0}^{\phi(f_2)}k_{f_1}(n-f_2,\beta) \le
         q^{s_2-f_2}\sum_{f_1=0}^{\phi(s_2)}k_{f_1}(n-s_2,\beta).
   \end{equation}
Indeed, using (\ref{eq:k3term}), we obtain
   \begin{align*}
    \sum_{f_1=0}^{\phi(f_2)}k_{f_1}(n-f_2,\beta)&=
 \sum_{f_1=0}^{\phi(f_2)}(k_{f_1}(n-f_2-1,\beta)+(q-1)k_{f_1-1}(n-f_2-1,\beta))\\
  &=k_{\phi(f_2)}(n-f_2-1,\beta)+q\sum_{f_1=0}^{\phi(f_2)-1}
       k_{f_1}(n-f_2-1,\beta).
   \end{align*}
Recall that $\phi(f_2+1)=\phi(f_2)$ or $\phi(f_2+1)=\phi(f_2)-1.$
In the former case,
  $$
  k_{\phi(f_2)}(n-f_2-1,\beta)+q\sum_{f_1=0}^{\phi(f_2)-1}
       k_{f_1}(n-f_2-1,\beta)\le q\sum_{f_1=0}^{\phi(f_2+1)}
              k_{f_1}(n-f_2-1,\beta);
  $$
in the latter,
  $k_{\phi(f_2)}(n-f_2-1,\beta)=k_{\phi(f_2+1)+1}(n-f_2-1,\beta)\le 0$
on account of (\ref{eq:beta}) and (\ref{eq:roots}). 
Repeating this procedure $s_2-f_2$ times, we arrive at
(\ref{eq:in}).

Note that $\phi(s_2)=s_1-1.$ Therefore
  $$
   U_L(a,0)\le q^{s_2}\sum_{f_2=0}^{s_2}k_{f_2}(n-\beta,\alpha)
    \sum_{f_1=0}^{s_1-1} k_{f_1}(n-s_2,\beta).
  $$
By (\ref{eq:cdk}), (\ref{eq:a}), and (\ref{eq:K0}) we have
  $$
    \sum_{f_1=0}^{s_1-1} k_{f_1}(n-s_2,\beta) =
     \frac{s_1 \binom{n-s_2}{s_1}(q-1)}{q\beta \binom{n-s_2}{s_1-1}}
        k_{s_1-1}(n-s_2,\beta),
  $$
  \begin{align*}
   \sum_{f_2=0}^{s_2} k_{f_2}(n',\alpha) &=
       \frac{(s_2+1)(k_{s_2+1}(n',0)
   k_{s_2}(n',\alpha)-k_{s_2+1}(n',\alpha)k_{s_2}(n',0))}
         {q\alpha k_{s_2}(n',0)}\\
  &=\frac{s_2+1}{q\alpha} k_{s_2}(n',\alpha)(W(0)-W(\alpha)),
  \end{align*}
where $n'=n-\beta$ and $W(x)=k_{s_2+1}(n',x)/k_{s_2}(n',x).$
Using these expressions, we can bound
$U_L(a,0)$ as
   $$
   U_L(a,0)\le \frac{(s_2+1)(n-s_2-s_1+1)(q-1)}
       {q^2\alpha\beta}K_s(a)(W(0)-W(\alpha)).
   $$
Hence using (\ref{eq:codebound}), (\ref{eq:F}), and (\ref{eq:f0}) we can write
  $$
   M\le v_s\frac{(s_2+1)(n-s_2-s_1+1)^2(q-1)^2 (\alpha + 2\beta)}{q^3\alpha^2\beta^2}
  \frac{(W(0)-W(\alpha))^2}{-W(\alpha)}.
  $$
Since $W(0)=(q-1)(n'-s_2)/(s_2+1)>0>W(\alpha)>-\infty$ as $\alpha$ ranges in
between the bounds in (\ref{eq:a}),
it is possible to find $\alpha$
such that $W(\alpha)=-W(0)$. With this choice and (\ref{eq:v}) the last
expression turns into (\ref{eq:fb2}).
The estimate (\ref{eq:ab2}) follows from (\ref{eq:arraybounds}).
\end{proof}

The proof of Theorem \ref{thm:2} is obtained by passing to asymptotics
in (\ref{eq:fb2}). Namely, let
  $
   n\to\infty, d/{nr}\to\delta , {s_1}/n\to\tau_1, 
    {s_2}/n\to\tau_2.
  $
By (\ref{eq:limrt}),
  $$
   \limsup_{n\to\infty} \frac\beta n=\gamma(\tau_1)(1-\tau_2),\quad
    \limsup_{n\to\infty} \frac \alpha n=\gamma(\tau_2)(1-\gamma(\tau_1)
     (1-\tau_2)).
  $$
Computing the logarithm on the right-hand side of (\ref{eq:fb2}), we observe
that the only term of exponential growth arises from $v_s$. 
Using standard estimates we obtain
   \begin{align*}
   \log_q v_s&=\log_q \binom n{s_1-1}\binom{n-s_1+1}{s_2}q^{s_2}(q-1)^{s_1
  +s_2-1}\\
      &\le n \big\{\tau_2+ h_q (\tau_1) + 
     (1-\tau_1) h_q \big(\frac{\tau_2}{1 - \tau_1})\big\}.
   \end{align*}
The tightest bound is obtained by computing the minimum of this expression
on $\tau_1,\tau_2$. The range of the variables $\tau_1, \tau_2$ is obtained
on observing that $n(q-1)/q^2$ and $n(q-1)/q$ are the 
maximizing values of $s_1,s_2$ for large $n$ (by a direct calculation
from the above expression; or, specializing from a general result in 
\cite{ros97}). The third restriction in the statement of the theorem
is implied by $\alpha+2\beta\le d.$
This completes the proof.   

\begin{figure}[t]\hspace*{-3.0cm}
      \scalebox{0.50}{\includegraphics[keepaspectratio]{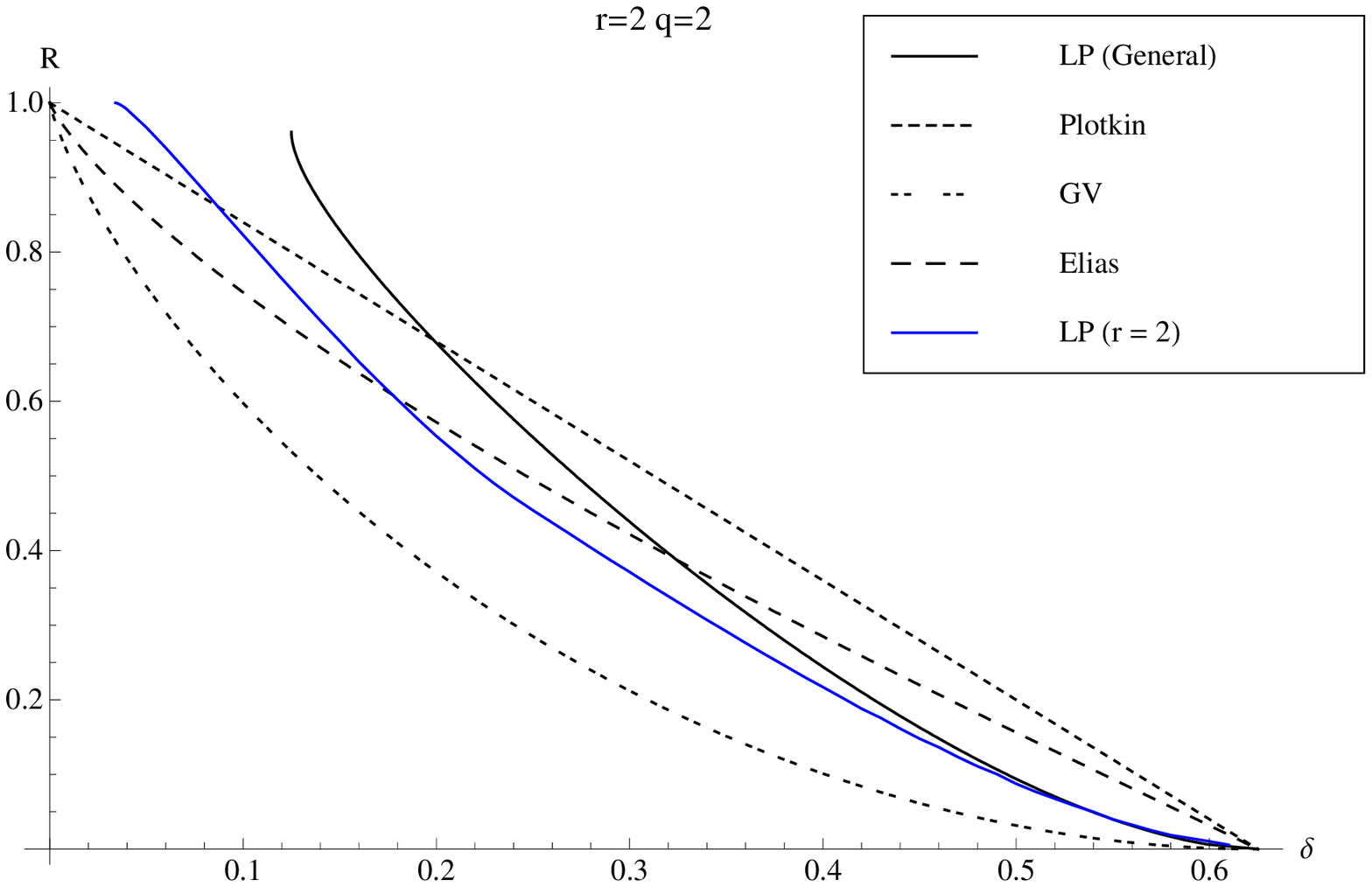}
\includegraphics[keepaspectratio]{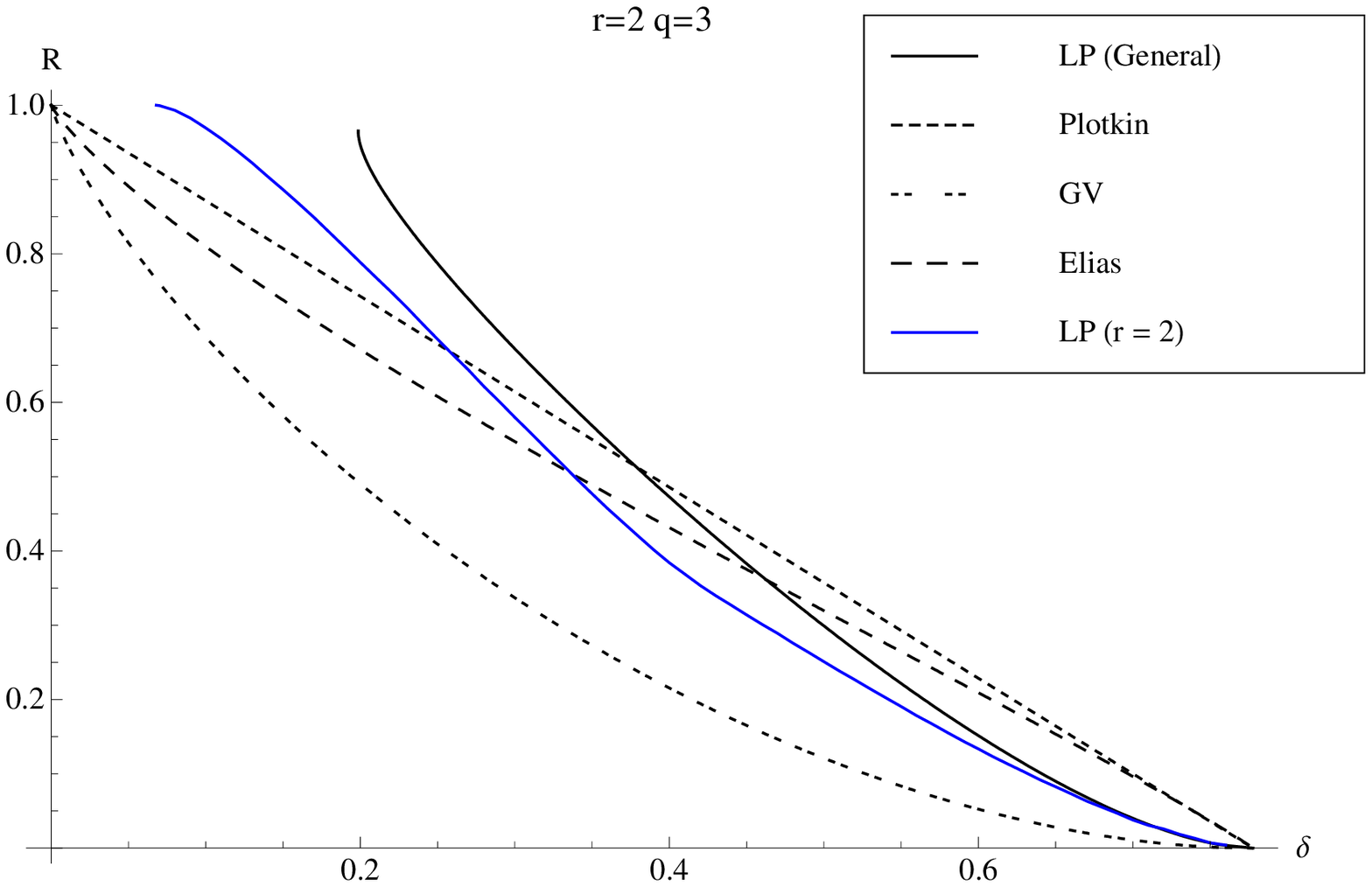}}\\
\hspace*{-3.0cm}
\scalebox{0.50}{\includegraphics[keepaspectratio]{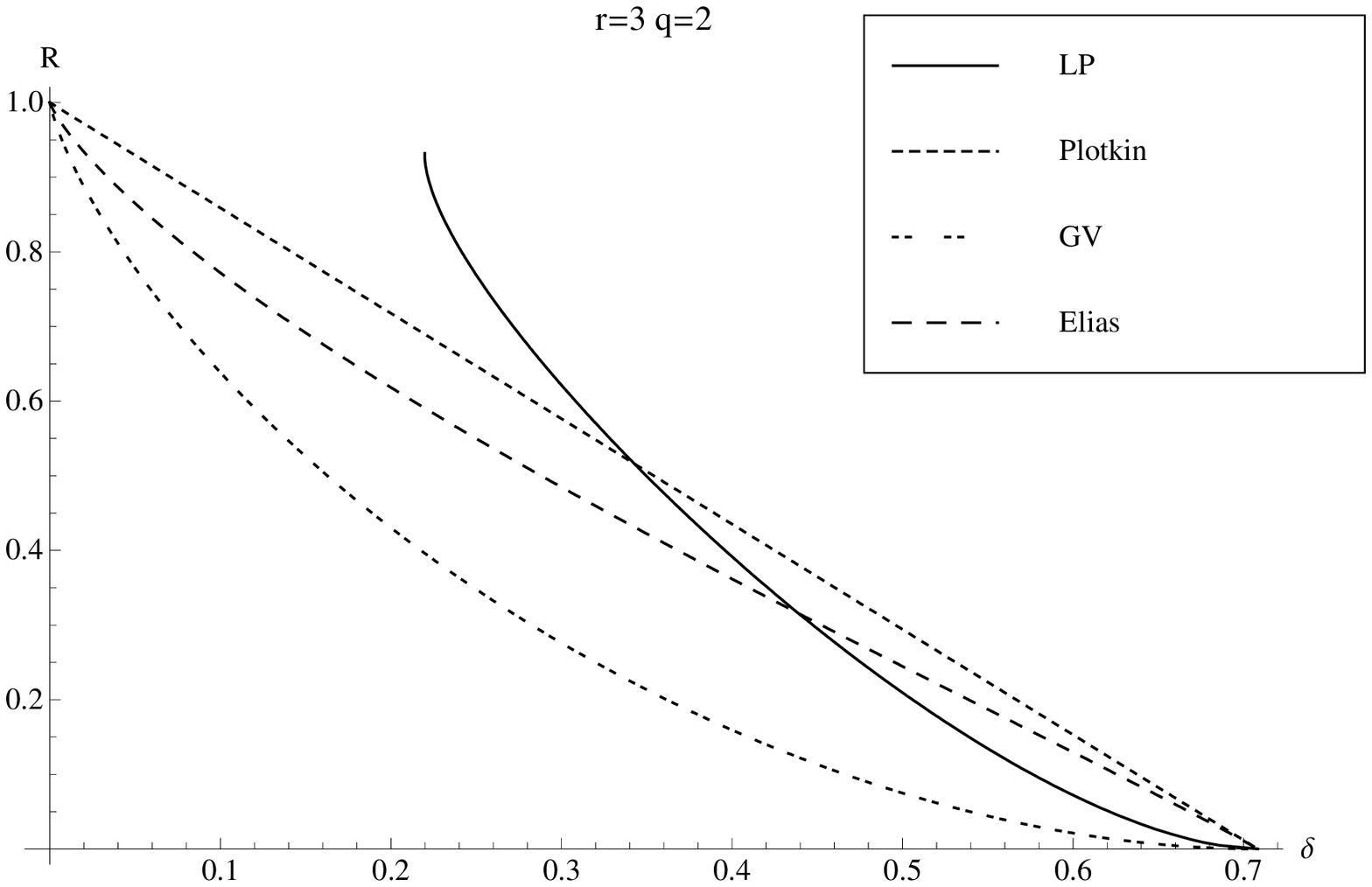}
\includegraphics[keepaspectratio]{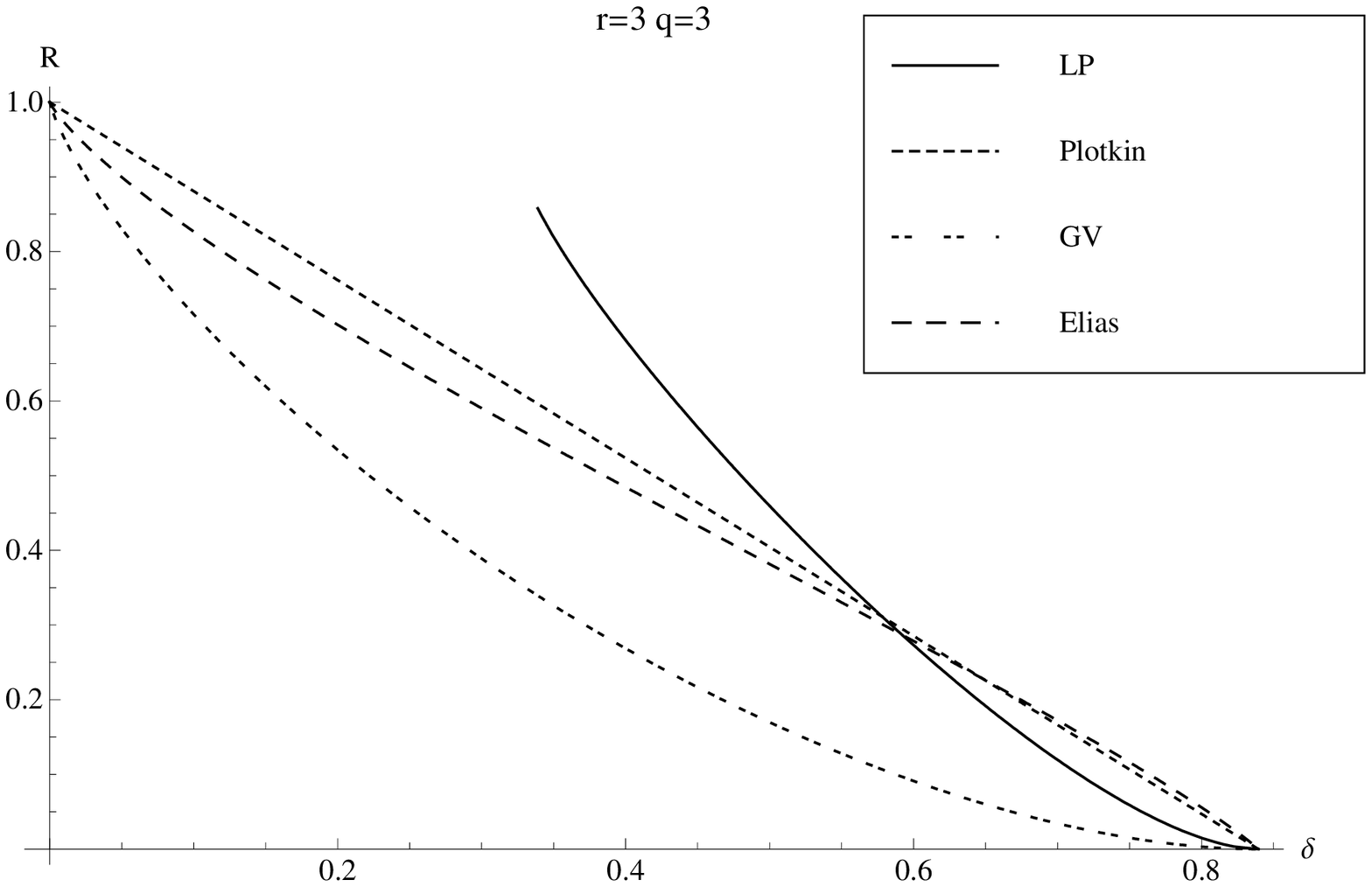}}
       \caption{Bounds for codes.}
       \label{l2v2} 
\end{figure}

Asymptotic bounds for ordered codes are shown in several plots in 
 Fig.~\ref{l2v2}.

\medskip{\em Remark.}
The NRT metric is an example of a wide class of metrics on the set
$\cQ^{N}$ termed poset metrics following Brualdi et al.~\cite{bru95}.
To define a poset metric, consider a partial order $P_\prec$ on the
set $[1,2,\dots,N]$. An {\em ideal} in the order is a subset closed under
the $\prec$ relation. The $P_\prec$-weight of a vector $\bfx\in \cQ^N$ is the
size of the smallest ideal that contains the nonzero entries of $\bfx.$
 For instance, for the NRT weight, the relation $\prec$ can be defined
as $(i_1,j_1)\prec(i_2,j_2)$ iff $i_1=i_2, j_1<j_2,$ where $i_1,i_2$ are the 
indices of the block. A {\em dual order} $P_\succ$ is formed of the same set of
chains as the order $P_\prec$ but with the signs reversed within each chain.
In \cite{mar99,skr01,dou02}, and in our paper, the poset duality is realized 
as the $C\subset \hr,C^\bot\subset \hl$ convention.
One of the main questions that arises in this context
is to characterise the association scheme that arises from the order
and in particular, to derive the MacWilliams-type relations. Partial 
orders that give rise to a univariate MacWilliams relation have been
described by Kim and Oh \cite{kim05}. On the other hand, rather little is
known about the multivariate case of which the NRT 
space is an instance.

\remove{
whether a given order supports a MacWilliams-type identity, or, in
other words, whether there exists a linear change of variables that
relates the weight enumerator $A_\prec$ of a linear code $C$ and
the weight enumerator $A_\succ$ of its dual code. For the NRT weight the
answer is provided by Theorem \ref{thm:mw} above.

A general criterion for the order to support a MacWilliams identity 
was given in Kim and Oh \cite{kim05}. Moreover, papers \cite{kim05} and 
Kim \cite{kim07} derived general MacWilliams relations
for a class of posets and weight enumerators. It is an interesting question
to study association schemes that arise from partial orders other than
the NRT order, to characterize explicitly their eigenvalues and to investigate
combinatorial properties of codes that arise in this way.}

\medskip{\sc Acknowledgment:} A.B. is grateful to William Martin for 
calling his attention to the problem of code bounds for the NRT space.
The authors are also grateful to M. Aaltonen for a useful discussion of his
work \cite{aal90}.

\def\cprime{$'$}
\providecommand{\bysame}{\leavevmode\hbox to3em{\hrulefill}\thinspace}

\end{document}